# Seismic events found by waveform cross correlation after the announced underground nuclear tests conducted by the DPRK on 12.02.2013 and 06.01.2016. Aftershocks or hidden nuclear tests?

Ivan O. Kitov, Dmitry I. Bobrov, and M.V. Rozhkov

**Abstract**
The multi-master method based on waveform cross-correlation has found several aftershocks of the third (DPRK3) and the fourth (DPRK4) announced underground nuclear tests conducted by the DPRK on 12.02.2013, and 06.01.2016, respectively. The second DPRK test had no reliable aftershock hypotheses at the level of the method sensitivity and resolution. The closest to the test site array station USRK was not operational before 20.12.2008 and the first DPRK test cannot be analyzed by the multi-master method based on waveform templates from two IMS stations - USRK and KSRS. The largest aftershocks of the DPRK3 and DPRK4 are close in magnitude to the first aftershock of the DPRK5 on 11.09.2016, and many aftershocks of the DPRK6. Three big aftershocks of the DPRK3 occurred on 25.05.2014, *i.e.* 467 days after the mainshock. The DPRK3 aftershock found at 6:43:03 UTC had a relative magnitude of 2.91. There were found two aftershocks of the DPRK4: on 19.02.2016 (0:28:07 UTC) and on 02.07.2016 (19:52:27 UTC). The largest aftershocks of the DPRK3 and DPRK4 could be interpreted as related to the cavity collapse process likely followed by a chimney collapse, not reaching the free surface, however, as no noble gases and/or radionuclide particles were detected. Seismic signals from all events found before the DPRK5 (09.09.2016), had a high level of cross-correlation with the waveform templates of the DRPK5 and DPRK6. These signals prove the shape similarity, and thus, the close spatial proximity of the seismic events after DPRK3/DPRK4 to the DPRK5/DPRK6 aftershocks. There were no other events found by the multi-master method since 01.01.2009, even when the highest resolution regime was used. The resolution and sensitivity of the multi-master method were tuned and tested using the data after 09.09.2016.

## 1. Introduction

The biggest underground nuclear explosions conducted by the DPRK were the last two tests on September 9, 2016, and September 3, 2017, with body wave magnitudes, $m_b$, of 5.09 and 6.07, respectively. Table 1 lists the principal parameters of all announced DPRK tests as estimated by the International Data Centre (IDC) of the Comprehensive Nuclear-Test-Ban Treaty Organization (CTBTO). The fifths explosion was the first event within the DPRK test site with automatically detected aftershock activity [Adushkin *et al.*, 2017]. The first aftershock on September 11, 2016, was so small that we were able to find it only using waveform cross-correlation (WCC) based on the signals from the previous, almost collocated, seismic events. Specifically, for this first aftershock, the signals from the DPRK4 were used as waveform templates. Automatic IDC processing missed this seismic event, but it was confirmed by analysts in the interactive analysis with the cross-correlation event hypothesis used as a seed event.

For waveform cross-correlation, WCC, between signals from the DPRK explosions, many seismic stations of the International Monitoring System (IMS) can be used. Column Nsta in Table 1 presents the number of stations with $m_b$ magnitudes used for the network magnitude estimation. The DPRK1 is the weakest explosion in the series ($m_b$=4.08±0.07). This explosion has no surface wave magnitude, Ms, since no LR-waves were detected and associated. The cross-correlation coefficients, *CC*, obtained from pair-wise cross-correlation of these explosions at IMS array stations are very high (*CC*>0.9) [Kitov *et al.*, 2014], and thus, can be used for the relative location with the accuracy of less than 100 m. The relative location of the DPRK explosions and the estimates of their depth of burial are two key parameters for the physical interpretation of the unusual characteristics of the DPRK aftershock sequence: the duration of more than five years and clustering near the DPRK5 and DPRK6 epicenters. The duration of five years is exceptional in comparison with the aftershocks sequence observed after nuclear explosions of the same size [Adushkin and Spivak, 1993]. These parameters are also important for the understanding of the DPRK3 and DPRK4 aftershocks.

The aftershocks of the DPRK underground explosions are small in magnitude and the smallest of them can be detected only at the two closest IMS arrays: USRK (distance ~410 km, DPRK4/station azimuth 35.8°) and KSRS (~440 km, 193.6°). There are several IMS and non-IMS 3-C seismic stations at regional, but they are not sensitive enough to detect the weakest aftershocks even with the matched filter method. Figure 1 presents a map with two IMS array stations USRK and KSRS and two non-IMS 3-C stations SEHB (South Korea) and MDJ (China), which detected the first aftershock on 11.09.2016 [Adushkin et al., 2017]. The DPRK test site is shown by a star. Instructively, these two non-IMS stations failed to detect the aftershock on 13.09.2016, which was smaller than the first DPRK5 aftershock.

Table 1. Six underground nuclear explosions conducted by the DPRK with parameters estimated by the IDC. $m_b$-ML difference increases with magnitude. The area of confidence ellipse, S, depends on the number of defining stations. The LR-waves from the DPRK1 are too weak to be detected and used for the Ms estimate.

| # | Date | Time | Lat, deg | Lon, deg | Nsta | $m_b$ | Ms | ML | S, km$^2$ |
|---|------|------|----------|----------|------|-------|----|----|-----------|
| 1 | Oct 9, 2006 | 01:35:28 | 41.31 | 129.02 | 16 | 4.08±0.07 | - | 3.89 | 880 |
| 2 | May 25, 2009 | 00:54:43 | 41.31 | 129.05 | 45 | 4.51±0.04 | 3.56±0.09 | 4.27 | 265 |
| 3 | Feb 12, 2013 | 02:57:51 | 41.30 | 129.07 | 66 | 4.92±0.04 | 3.95±0.05 | 4.52 | 180 |
| 4 | Jan 6, 2016 | 01:30:00 | 41.30 | 129.05 | 63 | 4.81±0.04 | 3.92±0.04 | 4.61 | 193 |
| 5 | Sep 9, 2016 | 00:30:01 | 41.30 | 129.05 | 74 | 5.09±0.04 | 4.17±0.05 | 4.29 | 153 |
| 6 | Sep 3, 2017 | 03:30:01 | 41.32 | 129.04 | 90 | 6.07±0.04 | 4.91±0.04 | 5.17 | 109 |

We have to use only the two closest IMS array stations to find the smallest DPRK aftershocks. In this case, the event definition criteria (EDC) adopted by the IDC for valid events in the Reviewed Event Bulletin (REB), which is the official IDC bulletin, are not applicable. The IDC EDCs require at least three IMS primary seismic stations to be associated with a valid event hypothesis [Coyne et al., 2012]. This requirement is based on the long-term statistics of the wrong association of automatic detections with false event hypotheses. The term "false event hypothesis" means that it was rejected by analysts as not matching the EDCs. The statistically acceptable level of event hypotheses reliability evenly distributed over the globe for a given configuration of IMS seismic stations can be reached only when the first P-wave arrivals at 3 primary stations are associated with a valid (*i.e.*, confirmed by an analyst) event hypothesis.

For the interactive analysis, an automatic Standard Event List (SEL3) is used as an input [Coyne et al., 2012]. This list may contain event hypotheses with two and even one (*e.g.*, weak regional events with several clear P- ($P_g$, $P_n$) and S- ($S_n$, $L_g$) phases) associated station. In the interactive phase, such events have to be improved to the level of 3 associated stations. Approximately a half of poor SEL3 events are rejected, but a few formally false events of interest with a high probability to be valid in physical terms are saved every day in the Late Event Bulletin (LEB) for the creation of long term statistics for further improvement in the IDC processing and products. There are many two-station events in the LEB, which are similar to the weak aftershocks found in this study. In several Flinn-Engdahl seismic/geographic regions, including the DPRK and surrounding territories, such two-station events are related to mining activity.

The WCC method is extremely powerful and can find signals deep in the ambient seismic noise [Adushkin et al. 2015]. No interactive analysis is feasible for such signals and one has to trust the result of cross-correlation. It is worth noting that WCC is similar to the discrete Fourier transform. Essentially, the coefficients of the Fourier transform are cross-correlation coefficients. For a harmonic with one frequency as a waveform template, the matched filter acts as the Fourier transform. The Fourier transform (spectrum) of the seismic signals generated by the DPRK aftershocks demonstrates that one can robustly estimate the amplitude of high-frequency harmonics with amplitudes much lower than the low-frequency noise level. In the same way, the matched filter method is able to find weak similar signals in the noise.

We have to develop a new set of event definition criteria based on detections obtained in automatic WCC processing in order to create reliable event hypotheses of the DPRK aftershocks using only two regional seismic stations. Comprehensive nuclear test monitoring implies a very high cost of one missed event of interest. For a given configuration of the IMS seismic network and available IDC analyst workforce, the task is to find all possible events in the data matching the adopted EDCs and confirmed by analysts. For the current study of the DPRK aftershock sequence, analysts are not able to confirm the event hypotheses using standard methods of interactive analysis - the signals from many aftershocks are close or below the ambient noise level. However, the weakest aftershocks are robust considering their cross-correlation with visible signals from aftershocks used as templates.

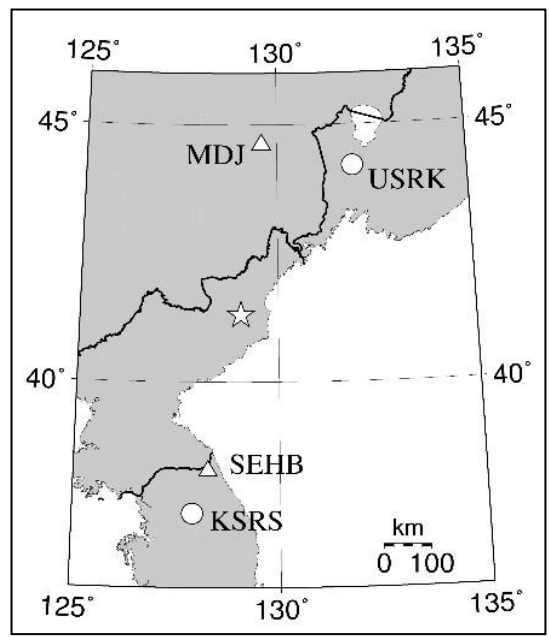

Figure 1. Positions of IMS stations USRK and KSRS relative to the DPRK test site shown by a star. Non-IMS stations SEHB and MDJ also detected the first reported DPRK aftershock on 11.09.2016.

Our approach to the event definition criteria is different from that adopted in nuclear test monitoring - we search for robust event hypotheses in the automatic regime and in some cases reject weak but potentially valid hypotheses during the days of extreme aftershock activity for the sake of consistent comparison of the events before and after the DPRK tests. The cost of missing a small aftershock after the test is low compared to a false positive event before the tests. In the latter case, one needs to explain the origin of the event demonstrating a relatively high cross-correlation with the DPRK aftershocks. To prove the presence of seismic activity not related to the DPRK explosion in the vicinity of the DPRK test site, one has to find a robust event hypothesis not related to the DPRK tests.

The DPRK-related data processing with the WCC method started in late 2014 with the DPRK3 as a master event, ME. The main target of WCC-based monitoring was to detect and estimate the relative location and magnitude of new DPRK explosions if any [Bobrov *et al*., 2017ab]. The failure to find aftershocks induced by the earlier, pre-DPRK5, tests is partially explained by the difference in seismic wave-field generated by explosions and earthquakes. We interpret the DPRK aftershocks as natural earthquakes in terms of source mechanism [Kitov *et al*., 2017]. An explosion-like source has a high-amplitude P-wave group. In relative terms, for the same seismic energy, earthquakes generate the P-waves of much lower amplitude, and the share waves, including the $L_g$-waves, are of much higher amplitude. This difference makes seismic signals from large explosions less effective in terms of the matched filter performance as a detector. Only with the aftershocks of the DPRK6 became it possible to effectively apply the matched filter detection to the induced seismicity. Currently, routine WCC processing of the DPRK-related data uses the multi-master method, which has higher resolution and finds much weaker aftershocks than at the initial stage.

The signals generated by earthquake-like events of the same or even lower magnitude are more effective than the signals from collocated explosions when used as waveform templates to detect similar earthquake-like events. The usage of waveform templates from the DPRK explosion was sub-optimal for the recovery of the DPRK aftershocks sequence, including the period before the DPRK5. Currently, almost ninety DPRK5 and DPRK6 aftershocks are available for an almost optimal setting of the matched filter detection of the aftershocks within the DPRK test site. The number of DPRK aftershocks is still growing as the process of interaction between the DPRK5 and DPRK6 chimney collapse is developing: there are more than a dozen aftershocks found in the first 9 months of 2021.

We have developed a multi-master cross-correlation approach optimizing the usage of the DPRK aftershocks and generating more robust event hypotheses [Bobrov *et al*., 2017c]. It is time to reprocess the whole period with data available at USRK and KSRS. In total, three aftershocks of the DPRK5 were found by cross-correlation before September 3, 2017, including the biggest one on September 11, 2016. The DPRK6 was the biggest explosion and initiated a long sequence of aftershocks observed even 4 years after the event (when this paper was under preparation), and likely will be observed in the near future. One cannot exclude a surface crater's creation as a result of the chimney collapse. The evolution of the aftershock sequence following the DPRK5 explosion as obtained by the multi-master cross-correlation method is described in [Kitov, 2021].

The period before 09.09.2016 is studied using the same multi-master cross-correlation method in order to find possible aftershocks of the DPRK2, DPRK3, and DPRK4. The DPRK1 was conducted on 09.10.2006 when IMS stations KSRS and USRK were not operational. Continuous data from station KSRS and USRK are available since 18.10.2008 and 20.12.2008, respectively. There are several tasks formulated for WCC processing of the historical data: 1) to process all data at USRK and KSRS since the start of USRK operation; 2) to find all aftershocks matching the predefined event definition criteria, EDCs, using routine processing parameters allowing fast calculation at a close to optimal resolution; 3) to retain the rate of the false event (as defined by the adopted EDCs) creation at a level of a few events per month; 4) to improve the solutions obtained at the routine stage with an extended set of control parameters providing the highest sensitivity and resolution; 5) process at least several months of USRK and KSRS data after DPRK2 through DPRK4 with an extended set of control parameters in order to find the smallest possible aftershocks related to cavity collapse. We expect to find all event hypotheses likely related to the aftershocks sequences of the DPRK2 through DPRK4 as well as possible natural events within the DPRK test site. At the same time, the rate of false event creation has to be under control in order to distinguish extremely weak valid and false events.

## 2. IDC raw data and metadata

We obtained seismic data for two IMS stations from the IDC database. This data set is available for any researcher through the virtual Data Exploitation Centre, vDEC. The raw data are used together with the REB, which contains the DPRK events parameters as well as station data, including the arrival times and SNR values for the first P-waves. In addition to the explosion data, Table 2 lists a few REB aftershock solutions and more solutions not formally matching the standard EDCs. These non-REB events are built by IDC analysts using two stations USRK and KSRS and are helpful for the WCC analysis, which is fully automatic.

Table 2. The events found by IDC analysts near the DPRK test site since DPRK1. Underground explosions are excluded. No weak events before the DPRK5 were reported. Alternative WCC solutions are listed for information, where available.

|   | IDC solution | | | | | | | WCC solution | | |
|---|---|---|---|---|---|---|---|---|---|---|
| # | Jdate | Date | hh:mm:ss | Lat | Lon | ML | Nsta[1] | hh:mm:ss | RM | Nass[2] |
| 1 | 2016255 | 11.09.2016 | 1:50:50 | 41.23 | 129.39 | 1.8 | 2 | 1:50:48 | 2.87 | 37 |
| 2 | 2017246 | 03.09.2017 | 3:38:32 | 41.32 | 129.06 | 3.2 | 12 | 3:38:30 | 3.81 | 37 |
| 3 | 2017266 | 23.09.2017 | 4:43:0 | 41.23 | 129.47 | 2.4 | 3 | 4:42:58 | 3.01 | 42 |
| 4 | 2017266 | 23.09.2018 | 8:29:16 | 41.23 | 129.35 | 3.0 | 4 | 8:29:14 | 3.61 | 49 |
| 5 | 2017285 | 12.10.2017 | 16:41:8 | 41.23 | 129.41 | 2.6 | 3 | 16:41:7 | 3.25 | 43 |
| 6 | 2017339 | 05.12.2017 | 14:40:53 | 41.37 | 129.01 | 2.6 | 3 | 14:40:50 | 3.14 | 40 |
| 7 | 2017343 | 09.12.2017 | 6:8:38 | 41.25 | 129.02 | 2.4 | 2 | 6:8:39 | 2.78 | 30 |
| 8 | 2017343 | 09.12.2017 | 6:13:33 | 41.34 | 129.15 | 2.9 | 3 | 6:13:32 | 3.42 | 35 |
| 9 | 2017343 | 09.12.2017 | 6:40:2 | 41.39 | 129.01 | 2.4 | 3 | 6:39:59 | 3.11 | 37 |
| 10 | 2018038 | 07.02.2018 | 21:46:2 | 41.30 | 129.17 | 2.6 | 3 | *21:46:0* | *3.34* | *16* |
| 11 | 2018038 | 07.02.2018 | 21:46:25 | 41.30 | 129.12 | 2.9 | 3 | 21:46:23 | 3.39 | 33 |
| 12 | 2018112 | 22.04.2018 | 19:25:11 | 41.29 | 129.33 | 2.0 | 2 | 19:25:9 | 2.71 | 47 |
| 13 | 2018112 | 22.04.2019 | 19:31:20 | 41.24 | 129.33 | 2.4 | 3 | 19:31:18 | 2.99 | 45 |
| 14 | 2019001 | 01.01.2019 | 22:20:28 | 41.29 | 129.29 | 2.8 | 2 | 22:20:27 | 3.19 | 35 |
| 15 | 2019079 | 20.03.2019 | 19:41:4 | 41.35 | 129.11 | 2.6 | 3 | 19:41:3 | 3.13 | 34 |
| 16 | 2019158 | 07.06.2019 | 5:18:41 | 41.30 | 129.13 | 2.7 | 3 | 5:18:39 | 3.20 | 47 |
| **17** | **2019180** | **29.06.2019** | **10:28:51** | **40.66** | **129.19** | **3.1** | **3** | | | |
| **18** | **2020210** | **28.07.2020** | **1:58:56** | **41.72** | **128.65** | **2.6** | **1** | | | |
| 19 | 2020220 | 07.08.2020 | 12:27:44 | 41.25 | 129.27 | 2.4 | 2 | 12:27:41 | 3.06 | 48 |
| **20** | **2020294** | **20.10.2020** | **22:22:18** | **41.82** | **128.05** | **2.4** | **3** | | | |
| 21 | 2020319 | 14.11.2020 | 14:45:39 | 41.19 | 129.54 | 2.5 | 2 | 14:45:35 | 3.00 | 38 |
| 22 | 2020319 | 14.11.2021 | 15:40:29 | 41.24 | 129.46 | 2.4 | 2 | 15:40:26 | 2.67 | 23 |
| 23 | 2020321 | 16.11.2020 | 18:10:30 | 41.25 | 129.42 | 3.0 | 3 | 18:10:28 | 3.35 | 24 |
| 24 | 2021030 | 30.01.2021 | 15:7:47 | 41.25 | 129.51 | 2.7 | 3 | 15:7:45 | 3.03 | 25 |
| 25 | 2021109 | 19.04.2021 | 6:48:51 | 41.21 | 129.41 | 3.0 | 3 | 6:48:48 | 3.49 | 46 |
| 26 | 2021164 | 13.06.2021 | 14:57:32 | 41.19 | 129.51 | 2.9 | 2 | 14:57:30 | 3.39 | 42 |
| 27 | 2021175 | 24.06.2024 | 17:9:34 | 41.25 | 129.21 | 2.0 | 2 | 17:9:32 | 2.65 | 43 |
| 28 | 2021209 | 28.07.2021 | 13:20:47 | 41.22 | 129.41 | 2.3 | 3 | 13:20:45 | 2.66 | 31 |

[1]Nsta - the number of stations with defining phases including non-primary IMS stations; [2]Nass - the number of associated templates in the multi-master method.

We do not visually inspect waveforms for signals found by cross-correlation. In some sense, we see only "spots of light on a radar screen" and hope that these spots correspond to real objects. In our case these are aftershocks. The reliability of such interpretation is based on the theoretical estimates of the matched filter resolution and sensitivity, comparison of the biggest aftershocks with the results of interactive analysis, and the broader experience of many researchers with the results of the WCC-based methods. It would be also helpful to have data or results from closer seismic stations.

In this study, we reprocess by the multi-master method the period between January 1, 2009, and September 9, 2016, for potential aftershocks and/or natural events in the close proximity of the DPRK test site. Seismic signals from such shallow natural events have to be similar to those from the DPRK aftershocks and also must have a close arrival time difference at stations KSRS and USRK, $tt_{diff}$. Using the DPRK explosions to estimate $tt_{diff}$, we obtain the value of 5.35 s for the DPRK6, 5.50 s for the DPRK5, 5.475 s for the DPRK4, 5.175 s for the DPRK3, and 5.25 s for the DPRK2. The accuracy of the onset time has to be high for signals with SNR>>20 and the difference between 5.3 s and 5.5 s is a conservative estimate of $tt_{diff}$ expected for the aftershocks of the DPRK5 and DPRK6. Any natural, *i.e.* not induced by any of the DPRK tests, seismic event close to the test site has to have $tt_{diff}$ close to those for the explosions. This is a very strong condition to reject events along the KSRS-USRK line. In the perpendicular direction, $tt_{diff}$ does not change much, despite the change in both travel times, and the events to east and west of the test site have the same $tt_{diff}$ as the explosions. The efficiency of the $tt_{diff}$ screening is well illustrated by the fact that several active mines within and around the DPRK are never confused with the DPRK aftershocks with the allowed deviation from the $tt_{diff}$ of ±1.5 s.

Table 2 includes the events within ~40 km around the DPRK test site. Almost all of the listed events are known aftershocks as obtained in standard IDC processing from the automatic SEL3 solutions and/or the WCC solutions as seed events. The final solutions obtained by the WCC method are also shown in Table 2 for a comparison. There are three LEB events highlighted in bold which are located at larger distances from the DPRK test site. Even though two of these three events are relatively large in terms of the number of associated IMS stations the WCC method did not find them. The reason for this failure is likely related to the fact that their source mechanisms (*e.g.*, mining blasts) and actual locations are far enough from those of the aftershocks. This observation shows that the WCC method is useful not only for the detection of relevant signals but also for the rejection of the signals not associated with the DPRK aftershock activity. Event #10 in Table 2 is confirmed by an analyst but the CC-based event hypothesis was rejected according to the EDCs adopted for the WCC method. A seed event obtained in automatic WCC processing is presented and highlighted bold-Italic. Its relative magnitude is too large for its cross-correlation properties. The most important conclusion we make from Table 2 for the purposes of this study is that there were no seismic events of any nature detected by the IDC around the DPRK test site before September 11, 2016, except the DPRK explosions excluded from this list.

## 3. Master events, waveform templates, and parameters obtained by waveform cross correlation

The WCC method allows significant improvements in magnitude estimation [Kitov *et al*., 2014, Bobrov *et al*., 2017a] and relative location of the events close is space [Bobrov *et al*., 2017b]. The DPRK tests represent an example of the relative location success because of the impulsive signals and large SNR values. The WCC method reaches the arrival time accuracy of a few thousandths of a second, and even less when re-sampling/spline interpolation is used, which corresponds to a few tens of meters for the $P_n$-velocity of 8 km/s, the travel time difference of 0.001 s corresponds to 8 m of the distance between events in the direction to the station.

The relative magnitude, d***RM***, is calculated as the logarithm of the ratio of L2-norms (lengths) of the signals from slave event, SE, │S│, and master event, ME, │M│. For the former, the length of the time window is the same as in the master, which found that signal. The set of frequency bands and lengths is predefined for a given master template and the length and frequency band with the highest SNR at the averaged CC-trace is selected for the relative magnitude calculation:

$$d\mathbf{RM} = \log(|S|/|M|) = \log|S| - \log|M| \tag{1}$$

Using an event as ME or SE may result in different d***RM*** absolute values because of a slight difference in the travel time residuals between channels of an array station for different locations of master events. This difference is small, however. The station d***RM*** estimates averaged over all available stations, ***RM***, provide an accurate and reliable estimate of the relative size of two similar events. Table 2 lists the relative magnitude, ***RM***, of the DPRK tests with the DPRK5 as a reference master event. The standard error of the mean value is also calculated in order to illustrate that the level of residual magnitude scattering over stations is much lower than by the standard method as listed in Table 1. The ***RM*** estimates are compared with those calculated by standard procedure from station magnitudes. The difference of the network magnitudes is small except that for the 2006 event – its magnitude was likely overestimated by the standard method.

This difference has a clear physical meaning for close events with similar signals. In addition to estimating the magnitude of the SE, the relative magnitude is a reliable dynamic parameter improving the reliability of phase association at several stations. The meaning of the dynamic matching of the arrivals is that the deviation of the station relative magnitude from the network should not be out of some narrow tolerance range. In routine WCC processing of the whole globe, the tolerance range is ±0.5 units of magnitude. In this study, the relative magnitude allows rejecting event hypotheses related to the side lobes of the WCC-detector allowing false detections from bigger seismic events located far from the corresponding ME. For example, the Tohoku earthquake generated a large number of seismic phases and produced several false events with magnitudes above 3.5. It was easy to reject them within the multi-master method approach as false by their large magnitude.

Table 3. Parameters of the DPRK tests obtained with waveform cross correlation.

| # | $m_b$ | RM (stdev) | dN, m | dE, m | Depth, m |
|---|---|---|---|---|---|
| 1 | 4.08 | 3.93±0.12 | -800 | 2400 | - |
| 2 | 4.51 | 4.54±0.06 | -450 | 100 | - |
| 3 | 4.92 | 4.96±0.07 | -850 | -220 | 1600 |
| 4 | 4.82 | 4.85±0.05 | -120 | -470 | 1350 |
| 5 | 5.09 | 5.09[1]±0.05 | 0 | 0 | 1300 |
| 6 | 6.07 | 6.11±0.03 | 0? | 1000? | 1650 |

[1]Reference event

For the accurate location of a SE relative to a ME with an estimated absolute location, one can use the set of signals with their arrival times obtained by the WCC method at each station j, $t_j$. For a SE collocated with a given ME, the travel time to the corresponding stations, $tt_j$, can be represented by the theoretical ME/station travel time or by the empirical travel time, *i.e.* by the difference of the arrival time of the signal from the ME at station *j* and the origin time of the ME. Then the origin time for the SE, $ot_j$ for station *j* can be estimated from the arrival time at station j:

$$ot_j = t_j - tt_j \qquad (2)$$

For J stations, the set of J arrival times is converted into a set of J origin times. (In the Local Association, LA, the process discussed in the next sections, we allow the origin time residual of 3 s.) In actual data processing, we suggest that the events sought by the WCC method are not collocated with the MEs and we have to estimate the best-fit position of the SE relative to the ME. For this, we introduce a grid of evenly distributed virtual epicenters. The grid size and spacing between its nodes are defined by the task. For two explosions spaced by a few hundreds of meters, a reasonable grid size is a few kilometers, and the spacing can be just a few tens of meters depending on the accuracy of the arrival time estimate by the WCC method.

For each node of a given virtual grid, the origin time at station j is calculated as a correction to the empirical travel time $t_j$. For $k$-th node the correction, $dt_k$, is calculated as a scalar product of the horizontal slowness vector, **s**, for corresponding seismic phase and the ME-node vector, **d**$_k$:

$$dt_k = \mathbf{s} \cdot \mathbf{d}_k \qquad (3)$$

The slowness is determined by the velocities of the corresponding seismic phase. We use standard values from the IASP91 velocity model. The more accurate values can be obtained from the regional travel time curves. Now, one can rewrite (3) in the form:

$$ot^k_j = t_j - tt_j + dt_k \qquad (4)$$

For all J stations, we move the virtual event location over the grid and calculate the average origin time and its standard error over all arrivals within ±3 s around the calculated origin time in each node. For the DPRK explosions, the observed travel (an equivalent of the origin) time residuals are within a few thousandths of a second. The point with the minimum standard deviation is considered as the relative location of the sought source. This node is the closest to the actual SE location.

Figure 2 presents the distribution of the RMS travel time residuals over the grid as obtained using four regional array stations USRK, KSRS, MJAR, and SONM for the DPRK4 as the ME and the DPRK5 as the SE. These 4 stations are characterized by different sampling rates: 20 Hz at KSRS, 40 Hz USRK, 50 Hz at SONM, and 80 Hz at MJAR. In order to make the sampling rate, and thus, the distance uncertainty associated with the discrete timing, uniform over stations as well as to improve the accuracy of the onset time estimation we resample all waveforms to 200 Hz. This rate provides the onset time uncertainty of 0.0025 s, which is equivalent to the source-station distance uncertainty of ~20 m for the $P_n$-wave with 8 km/s.

Having 4 stations, one can determine the relative location with a 10 m to 20 m uncertainty. For poor SE/ME signals (*e.g.*, the signals from the DPRK aftershocks), however, the uncertainty increases since the onset time can be biased by the ambient microseismic noise and the difference in seismic phases generation by sources of different nature, *e.g.*, underground explosions, natural earthquakes, aftershocks, underground collapses, and mining blasts. For the DPRK underground tests, we have 60 SE/ME pairs for relative location. The number of stations associated with the DPRK explosions changes with magnitude and new stations added to the IMS seismic network with time since the DPRK1. All qualified stations with SNR of the associated signals above the predefined quality threshold (say, SNR=10) are used in the relative location. We also use regional and teleseismic networks separately and jointly. Using the pair-wise relative locations, one can calculate the joint relative locations minimizing the overall standard error. Unfortunately, the DPRK6 is excluded from the joint relative location due to its finite size disturbing the travel time predictions. Figure 3 presents the joint relative locations of the five events. For the purposes of this study, the location of the DPRK2 through DPRK4 relative to the DPRK5 (and DPRK6) with known aftershocks is important.

Table 3 lists the distance in meters to the north, dN, and east, dE, relative to the DPRK5. One can see that the DRPK5 is to the north from all events, except likely DPRK6, which position cannot be determined with the accuracy required for the relative location. The DPRK6 position is most likely within 1.0 km to 1.5 km to the west of the DPRK5. The distance between DPRK4 and DPRK5 is approximately 550 m, which is a reasonable value for the events with body wave magnitudes around 5. A conservative estimate of the yield corresponding to $m_b$=5.0 explosion in hard rock is 25 *kt*. For the cube-root elastic radius scaling and 100 $m/kt^{1/3}$ empirical regression coefficient, one obtains the DPRK3 and DPRK4 elastic radius of approximately 200 m to 250 m, *i.e.* they are beyond the elastic radii of each other, and thus, the full containment is not at risk. For the DPRK5 and DPRK6, the elastic radius is 300 m and 700 m, respectively. They should not be closer than 1 km to each other for the full containment.

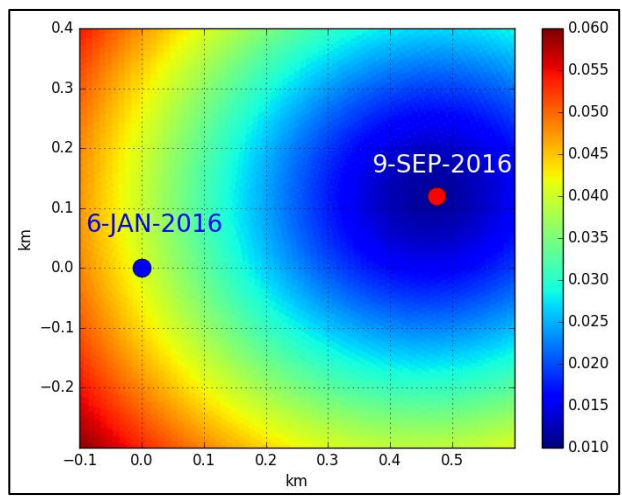

Figure 2. Example of the relative location: the position of the DPRK5 event (dot in centre of blue circles) relative to the DPRK4. The distribution of the RMS travel time residuals is shown by colour. Scale in seconds.

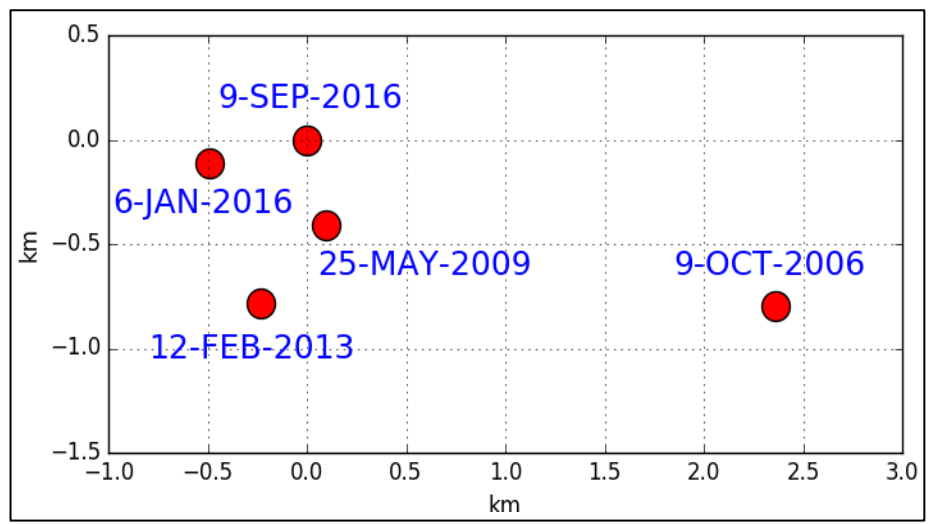

Figure 3. Relative location of the first five DPRK test as obtained by cross correlation at IMS regional stations. The DPRK6 test is not well located in relative sense because of its size (elastic radius of approximately 1 km in radius) destroying the point source approach to the relative location.

The magnitude effect on the accuracy of relative location is the result of the elastic source size - seismic wave emission may depend on event-station azimuth when the source is larger than the wavelength of the wave used for detection and the speed of source expansion is larger than that of the P-wave. For example, the DRPK6 elastic radius is approximately 1 km and the shock wave is faster than the P-wave in the embedded rock. The travel time to USRK and KSRS, which are practically in opposite directions to the explosion, should both be smaller than from a point source in the hypocenter of the explosion. Then the travel time difference at KSRS and USRK will not match the prediction obtained as a correction of the travel time difference for the master event, *e.g.* the DPRK5. This makes the relative location accurate only to ~1 km compared to 100 m for the other 5 events.

The DPRK3 is almost due south of the DPRK5 at approximately 900 m. This position makes the interaction of the corresponding cavities negligible. The DPRK4 is closer to the DPRK5. The distance between potential hypocenters of the DPRK3/DPRK4 aftershocks and the DPRK5/DPRK6 aftershocks is compatible. Therefore, the WCC method is applicable to the potential DPRK3 and DPRK4 using the DPRK5 and DPRK6 aftershocks as master events with their associated waveform templates.

The relative locations in Table 3 can be compared with the distance estimates following from the $tt_{diff}$ (the difference of arrival times measured at KSRS and USRK) measurements for the DPRK tests. For example, the $tt_{diff}$ for the DPRK5 is 5.5 s and 5.175 s for the DPRK3. The difference in $tt_{diff}$ of 0.325 s has to be divided by 2 for distance calculations: 0.1625[s]·8000[m/s]=1300 m. The relative location in Table 3 is approximately 1 km. It is obtained from the relative travel time differences for the DPRK5 and DPRK3 at many IMS stations, which are more accurately estimated by cross-correlation of the high-SNR impulsive signals. The difference of 0.4 km between these two estimates of the distance between the DPRK5 and DRPK3, is within the uncertainty related to the accuracy of the onset time of 0.05 s to 0.1 s in the interactive analysis. The onset time uncertainty of 0.05 s corresponds to 400 m for the $P_n$-velocity of 8 km/s. The effect of $P_n$-wave velocity variation in the range 7.7 km/s 8.3 km/s on the distance estimate is rather small. The distance between DPRK6 and DPRK5 epicenters along the KSRS-USRK line for the difference in the $tt_{diff}/2$ estimates of 0.075 s ± 0.05 is 500 m ± 400 m. The DPRK6 is likely located a few hundred of meters to the south of the DPRK5 and 700 m to 1500 m to the east. The distance in the E-W direction does not affect $tt_{diff}$.

The depth of burial is also a key parameter for the understanding of the aftershock activity. The DPRK tests were conducted in horizontal tunnels in a hill with linear sizes of several kilometers. The high-resolution satellite images reveal a possible entrance portal, but do not reveal the structure of underground tunnels. The relative location of the DPRK epicenters is not helpful for the estimation of their depths. We estimate the depth of burial (more accurately, the distance to the shock wave reflection at the free surface of the hill) using the method of Tape and Tape [2015] as based on a uniform discretization of the moment tensor space with the following determination of the optimal moment tensor and depth by comparing observed seismograms with synthetic waveforms. This comparison is based on waveform cross-correlation. Figure 4 presents the preliminary results of the moment tensor estimation as a function of source depth for the DPRK4 and DPRK5 [Bobrov *et al.*, 2017]. Teleseismic P waves are windowed around an analyst-picked first arrival filtered at 0.6 Hz to 4.5 Hz. Regional Rayleigh waves on the vertical component are selected in a 180-second window filtered 10-50 s period. Instrument responses are removed from observed signals prior to processing. Teleseismic Green's functions are calculated with the stationary phase approximation using the AK135 velocity model for the propagation path and CRUST1.0 for the source and receiver locations, and the regional surface waves are synthesized with the wave-number integration, using Robert Herrmann's code (CPS package). The optimal solution shown as beach-ball at joint PDF/Depth plot corresponds to 1.3 km for January 6, 2016 event and 1.0 km for the September 9, 2016 event, with pre-dominant positive isotropic source (61% and 51%), accompanied by well-pronounced DC component (26% and 38%) and 13% and 11% CLVD.

The final depth estimates for the DPRK3 through DPRK6 are given in Table 3. They are slightly different from the preliminary results. More efforts are needed to improve the accuracy and to reduce the uncertainty of the DPRK depth estimates. For the aftershocks study, the depth of burial provides important information on the possible mechanical effects of explosions. The scaling law for underground nuclear tests suggests that the optimal depth of burial in hard rock is approximately equal to the elastic radius of the explosion with a given yield. For a 1 *kt* event, the depth is ~100 m. For a 1*Mt* explosion, the depth of burial has to be $100[m]*(1000)^{1/3} = 1000$ m, i.e. ten times larger than for a 1 *kt* explosion. According to the scaling law, the DPRK events were over-buried, and thus, the absence of radioactive traces (noble gases and particles) in the atmosphere has a clear physical explanation. At the same time, the gravitational energy related to the cavity creation increases approximately as a linear function of the depth of burial, with the cavity volume decreasing with depth according to the scaling law, however.

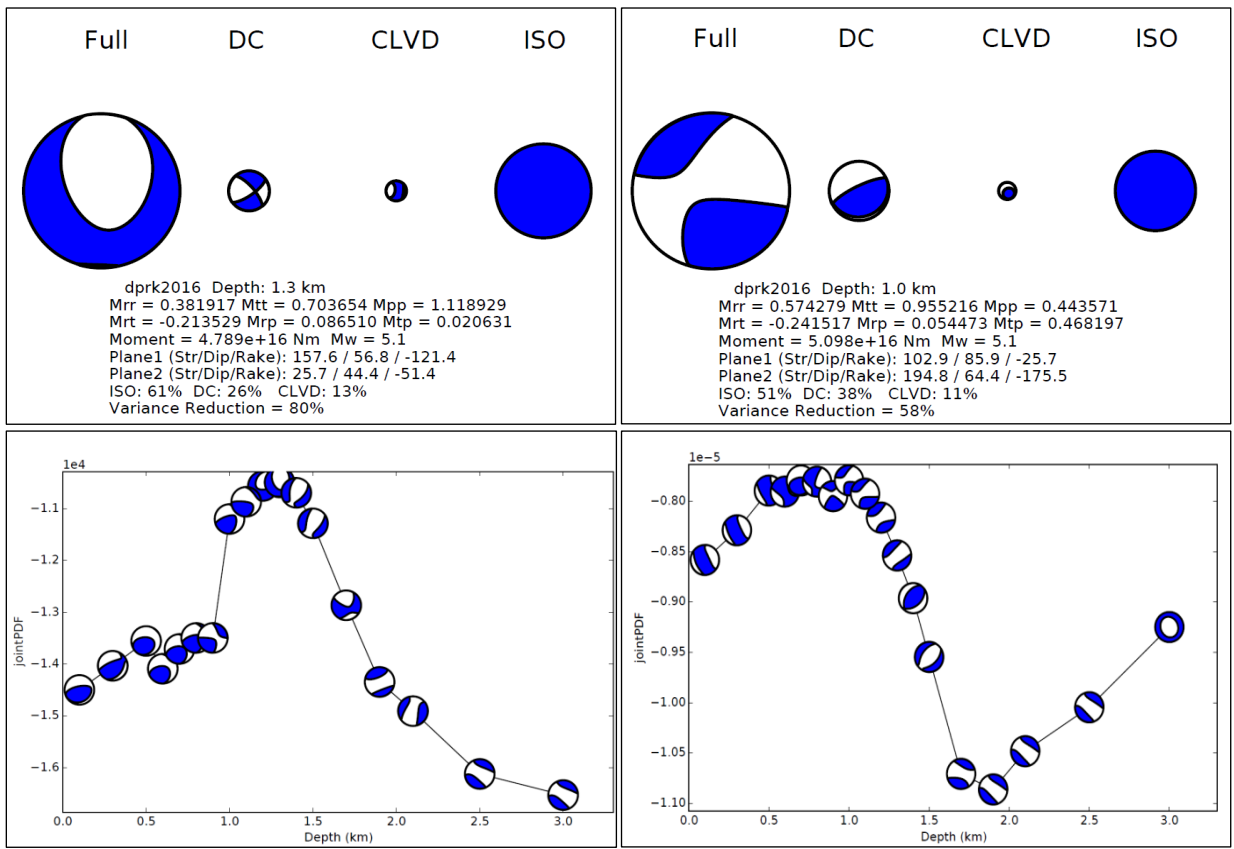

Figure 4. The depth of burial estimates for the DPRK4 (left panel) and DPRK5 (right panel). Simplistic representation of the fit (joint PDF) between observed and predicted seismograms depending on depth and moment tensor demonstrates both the most probable depth and the uncertainty of the depth estimate. For the DPRK5, there is a wide range of almost the same PDF level.

The distance between the DPRK5 and DPRK6, their magnitudes and depths of burial created a unique configuration for an extended aftershock sequence observed 5 years after the DPRK5. We also reported a clear clustering of the aftershocks belonging to the DPRK5 and DPRK6 sequences as described by their better cross-correlation within the clusters and much lower correlation between the clusters [Adushkin *et al*.,2021; Kitov, 2021]. The DPRK5 and DPRK6 aftershocks also cluster in time creating an alternating pattern likely corresponding to the interaction between the PDRK5 and DPRK6 collapse chimneys. The DPRK4 aftershocks are likely closer in space to the DPRK5 and it is important to estimate their similarity to both clusters. The DPRK3 is far enough from both clusters and may be similar to all waveform templates.

## 4. Cross correlation, waveform templates, detection

Calculation of cross correlation coefficient for a given discrete waveform and a waveform template follows standard definition:

$$CC(t_i) = m(\tau_n) \cdot u(t_{i+n}) / (\|m(\tau_n)\| \cdot \|u(t_{i+n})\|) \qquad (5)$$

where $CC(t_i)$ is the cross-correlation coefficient for a discrete time $t_i$, $m(\tau_n)$ is the waveform template of the length N, $u(t_i)$ is the original trace of the length I≥N, $\|\cdot\|$ determines the L2-norm for the corresponding time series of length N. The waveform template is taken as a signal at the same station with the start point $t_0$, where N·d$t$ + $t_0$ < $t_i$, and d$t$ is the time step. For an array or 3-C station, the aggregate correlation the coefficient is calculated by averaging over K channels:

$$CC(t_i) = [\Sigma_k CC_k(t_i)]/K \qquad (6)$$

where $CC_k(t_i)$ is the cross-correlation coefficient at time $t_i$ for channel $k$. It is important that for an array station no time shifts between channels are applied when calculating $CC(t_i)$. This is due to the fact that the template and sought signal are supposed to be synchronized at all channels as a result of the presumed spatial closeness of the ME (template) and SE. One can formally apply beam-forming to the CC-traces, however, and obtain an F-K plot, which is a standard representation for beam-forming. Deviation from the empirical vector slowness of the master event can be the result of the difference in locations of the master and SE, the difference in their source functions, and the effect of the ambient noise. The latter reason is prevailing for weak signals, especially for those hidden in the noise.

The CC-trace was obtained with the original unfiltered data where low-frequency noise is dominating. To improve the $CC$ estimates we apply filtering of the original waveforms in the frequency range where the signal spectra of aftershocks are above the ambient noise spectral level. Five 4-order Butterworth band-pass filters are introduced: 1.0 Hz to 2.0 Hz, 1.5 Hz to 3.0 Hz, 2.0 Hz to 4.0 Hz, 3.0 Hz to 6.0 Hz, 4.0 Hz to 8.0 Hz. Using the best signals from the DPRK explosions and aftershocks, fixed sets of multi-channel waveform templates for stations USRK and KSRS were prepared for each of the five filters. The total length of all templates is 205 s, and includes a 5 s lead before the signal arrival time at the central element of an array and 200 s of the signal. The lead allows accommodating possible arrival mistiming as well as slightly earlier arrivals at array elements, which are closer to the source than the central station.

A template of 200 s includes all regular regional seismic phases, from $P_n$ to $R_g$. Figure 4 presents five templates at channel KS01 of station KSRS corresponding to the five filters as obtained for the DPRK5. Instead of time, we use counts since the number of points in the template time series is more important for cross-correlation than formal time intervals. Table 4 lists 23 aftershocks used as a master event [Kitov *et al*., 2018; Kitov, 2021]. The full length of the template for KSRS with a 20 Hz sampling rate is 4101 count. There are several regular phases present in the template: $P_n$, $P_g$, and $L_g$, which are also identified in the REB. The relative amplitudes of P- and S-phases depend on frequency band: $L_g$ is a prominent phase for the two low-frequency filters and then fades away with increasing frequency. In turn, the $P_n$ and $P_g$ phases are dominating at higher frequencies. As a result, the highest cross-correlation coefficient is obtained when a short correlation window and higher frequencies are used.

The difference between neighboring channels can be higher than the difference between signals from two spatially close events at the same channel. Figure 7 depicts templates at four different channels of USRK from the same aftershock. All templates correspond to filter #3 which is efficient for weak signals from the DPRK aftershocks. The visual difference between channels is significant and can be explained by the dispersion and complexity of the $L_g$-wave. For USRK with a 40 Hz sampling rate, the template length is 8201 count.

We use a multi-length approach for cross-correlation windows, from 20 s to 120 s with a 20 s increment in order to address the possibility of different seismic phases to define the level of similarity: the DPRK nuclear tests have relatively short but high-amplitude $P_n$-phase and $P_g$-phase but a relatively low-amplitude and lengthy $L_g$-wave. The DPRK aftershocks usually have only the $L_g$-phase is visible and the $P_n$-phase is very close in amplitude to the ambient seismic noise. The changing cross-correlation window length, CLW, allows exercising the similarity of various parts of the signals from the DPRK explosions and aftershocks and to partially avoid noise influence.

Table 4. Parameters of 23 DPRK aftershocks used as master events. Aftershocks from #2 to #11 belong to the DPRK6 cluster and from #12 to #23 are the DPRK5 aftershocks [Kitov, 2021]. #1 is the largest aftershock occurred 8.5 min after the DPRK6.

| N | Jdate | Time | # | NassR | NassF | NassKSRS | NassUSRK | RM |
|---|---|---|---|---|---|---|---|---|
| 1 | 2017246 | 03:38:30 | 1 | 36 | 37 | 17 | 20 | 3.81 |
| 2 | 2017266 | 04:42:58 | 1 | 42 | 42 | 21 | 21 | 3.01 |
| 3 | 2017266 | 08:29:14 | 2 | 48 | 49 | 25 | 24 | 3.61 |
| 4 | 2017285 | 16:41:07 | 1 | 43 | 43 | 18 | 22 | 3.25 |
| 5 | 2017304 | 10:20:12 | 1 | 18 | 20 | 5 | 15 | 2.46 |
| 6 | 2017335 | 22:45:54 | 1 | 32 | 35 | 18 | 17 | 2.90 |
| 7 | 2017339 | 14:40:50 | 1 | 42 | 40 | 19 | 21 | 3.14 |
| 8 | 2017340 | 16:20:04 | 1 | 23 | 31 | 14 | 17 | 2.61 |
| 9 | 2017343 | 06:08:39 | 1 | 22 | 30 | 11 | 19 | 2.78 |
| 10 | 2017343 | 06:13:32 | 2 | 41 | 35 | 17 | 18 | 3.42 |
| 11 | 2017343 | 06:39:59 | 3 | 39 | 37 | 18 | 19 | 3.11 |
| 12 | 2016255 | 01:50:48 | 1 | 36 | 37 | 19 | 18 | 2.87 |
| 13 | 2017246 | 09:31:28 | 2 | 31 | 41 | 19 | 22 | 2.67 |
| 14 | 2018036 | 10:32:30 | 1 | 36 | 36 | 12 | 24 | 2.66 |
| 15 | 2018036 | 20:07:29 | 2 | 28 | 37 | 14 | 23 | 2.76 |
| 16 | 2018036 | 21:57:35 | 3 | 23 | 27 | 8 | 19 | 2.75 |
| 17 | 2018037 | 04:49:36 | 1 | 22 | 24 | 5 | 19 | 2.64 |
| 18 | 2018037 | 10:12:30 | 2 | 17 | 25 | 5 | 20 | 2.59 |
| 19 | 2018037 | 10:53:52 | 3 | 49 | 47 | 23 | 24 | 3.02 |
| 20 | 2018038 | 21:46:23 | 1 | 47 | 33 | 17 | 16 | 3.38 |
| 21 | 2018039 | 17:39:17 | 1 | 41 | 42 | 19 | 23 | 2.74 |
| 22 | 2018112 | 19:25:09 | 1 | 36 | 47 | 22 | 25 | 2.71 |
| 23 | 2018112 | 19:31:18 | 2 | 50 | 45 | 25 | 20 | 2.99 |

# - order number of an aftershock in a given day. The master event identifier includes the date and the order number, *i.e.* 2017266_2 - the second aftershock on 2017266. NassR - the number of associated templates in routine processing. NassF - the number of associated templates in final processing. NassKSRS and NassUSRK - the number of templates associated for stations KSRS and USRK, respectively.

By design, the cross-correlation detection procedure is similar to standard IDC detection [Coyne *et al*., 2012]. The first step is to filter the original multi-channel raw waveforms using one of the filters predefined in the template creation procedure. Then, for each individual channel of an array or 3-C station, we calculate a time series of cross-correlation coefficients, *CC*, using the permanent waveform templates at corresponding channels. The CC-traces are averaged over all channels of the array/3-C station. There are no time shifts between the array channels since all CC-traces have to be synchronized in time by the requirement of almost perfect co-location of the MEs and SEs.

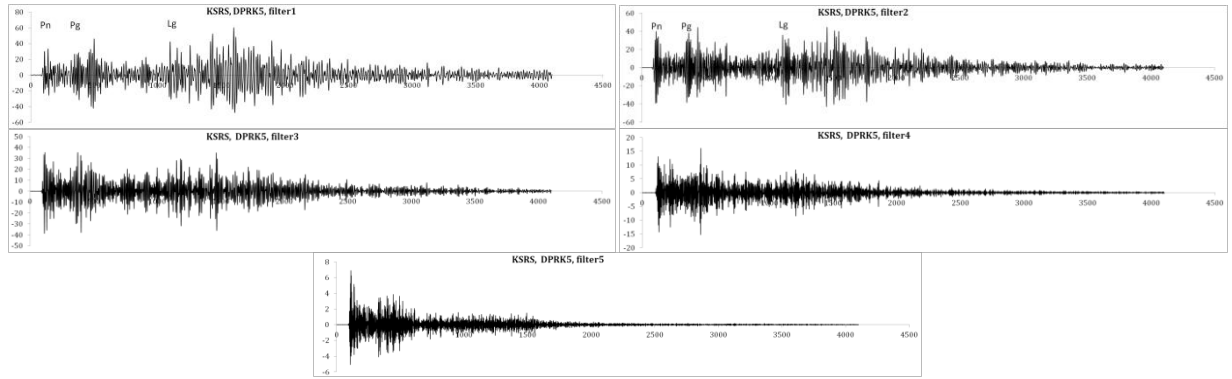

Figure 4. Waveform templates of the DPRK5 explosion at IMS array station KSRS (channel KS01) filtered with filters #1 through #5. The full template length is 4101 counts or 205 s (sampling rate of 20 Hz), including 5 s before the arrival time at the central element of the array. We use counts instead of time since cross-correlation is defined by the number of samples in the template. The first ($P_n$- and $P_g$-wave) arrivals are dominating over other seismic phases at higher frequencies and the amplitudes of the $L_g$-wave and $P_n$-wave are compatible when filters #1 and #2 are applied.

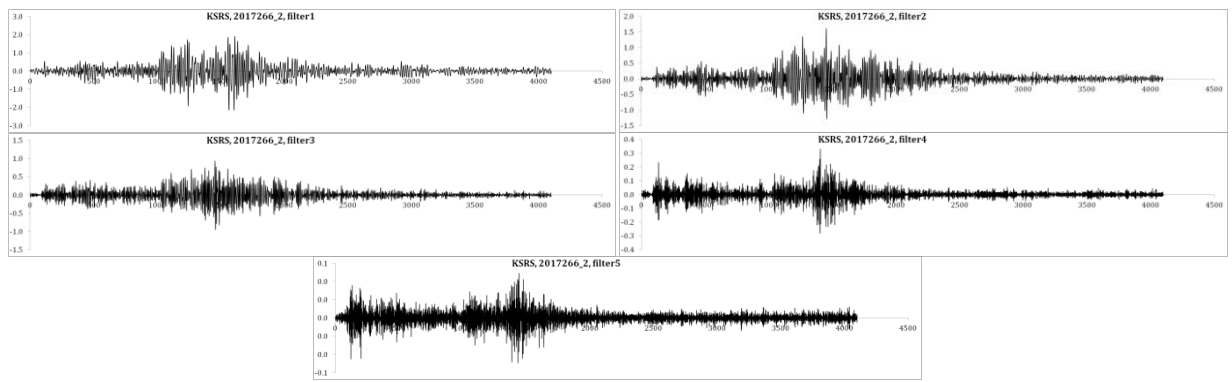

Figure 5. Templates of the second aftershock observed on September 23, 2017 (2017266_2) at IMS array station KSRS (channel KS01) filtered with filters #1 through #5. The $L_g$-wave is the most prominent at lower frequencies. The $P_n$-wave is seen at the higher frequencies. The P-to-$L_g$ amplitude ratio is different at higher and lower frequencies for explosions and aftershocks [Kitov and Rozhkov, 2017; Kitov *et al*., 2018]. This makes templates of explosions less effective in weak aftershocks finding.

For the autocorrelation or close to autocorrelation (*e.g*., the DPRK explosions with *CC*~1.0) cases, the peak *CC*-value at the average CC-trace provides an extremely accurate arrival time estimate. When the sought signal has a low amplitude (or the ambient seismic noise has a large amplitude) and it is different from the template, many local maxima are observed and it is not easy to select the proper peak and to estimate the arrival time. To accommodate this possibility into the CC-detection procedure, we use SNRcc=STA/LTA, where STA is the Short-Term Average and the LTA is the Long-Term Average. The STA and LTA have many definitions and we use the absolute average instead of a more common approach using RMS values. The SNRcc is calculated at the average CC-trace. This procedure is repeated for all filters and CWLs. For a given time count, the maximum SNRcc value among all filter/window length combinations is saved as the final SNRcc. Therefore, the SNRcc time series consists of the maximum SNRcc values among all combinations of filters and CWLs as the standard SNR time series consists of the maximum SNR values among all filters and beams [Coyne *et al*., 2012].

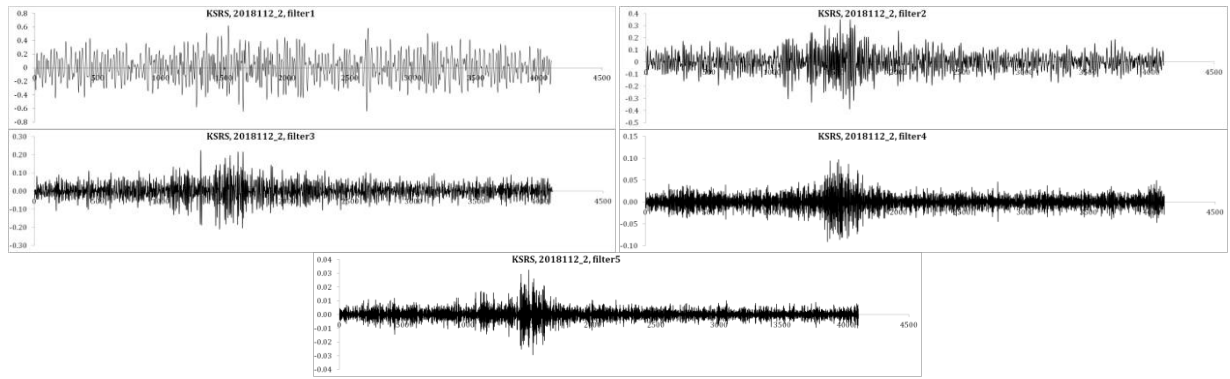

Figure 6. Templates of the second aftershock observed on April 22, 2018 (2018112_2) at KSRS (channel KS01) filtered with filters #1 though #5. There is no clear $P_n$-wave arrival and the $L_g$-wave is seen for filters 2 through 5.

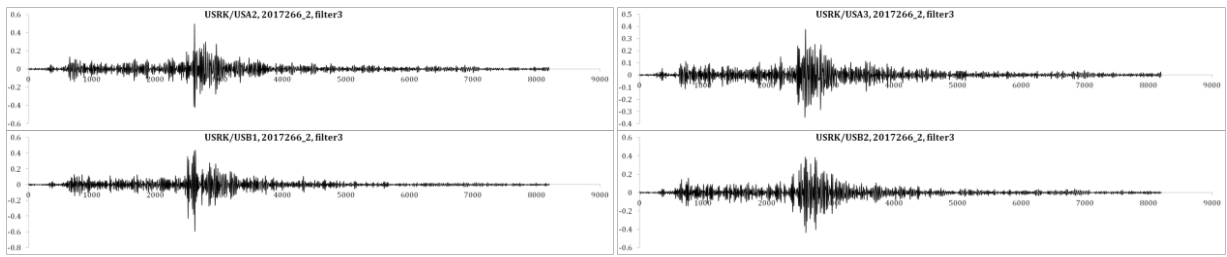

Figure 7. Templates of the second aftershock observed on September 23, 2017 at four different channels (USA2, USA3, USB2, USB2) of array station USRK. The difference between waveforms is significant.

The cross-correlation coefficient is a nonlinear function of the original waveform: a signal of any amplitude and length is converted into a single value between -1.0 and +1.0. Figure 8 displays a CC-trace for the case of DPKR6 signal correlation with its own template at KSRS. This is autocorrelation with the *CC*=1 as the upper left panel demonstrates. There is another signal clearly seen 8.5 min later as related to the biggest DPRK6 aftershock. These two events are practically collocated and *CC*≈0.6. The SNRcc trace shown in the left lower panel demonstrates the effect of the pre-signal noise. For the DPRK6 signal, SNRcc≈90, and for the aftershock is SNRcc≈30, *i.e.* the difference in SNRcc is much larger than the difference in *CC*. The noise effect is related to the presence of high-amplitude signals generated by the mainshock before the aftershock signal. As a result, the noise is not random and the *CC* is much larger before the aftershock signal as well as the LTA level. This effect suppresses the SNRcc estimates for the aftershock relative to the case when no coherent signals exist in the ambient noise.

For signal detection, there is another problem associated with the nonlinear conversion of an original seismic trace into a cross-correlation trace – a much larger uncertainty in the arrival time estimate. A physical signal of *T* seconds started at time $t_0$, which is similar to some template of the same length, by definition has to be more coherent with the template than with the ambient noise. As a result, one can find that the CC-traces start to reveal the presence of the sought signal almost the same time as the tail of the template touches the signal, at time $t_0$-*T*, *i.e.* long before the physical signal arrival time. This effect is observed due to the signal/template coherence and the absence of coherence between the template and the ambient seismic noise. When the template meets the sought signal, the *CC*-trace starts to increase in amplitude above the *CC*-trace obtained for the ambient seismic noise. Figure 9 zooms in the autocorrelation case: two intervals are shown: 250 s and 5 s. The CC-trace starts to grow in amplitude and the SNRcc reaches the detection threshold ~50 s before the physical signal demonstrating the intrinsic difference between physical and CC signals. The peaks of CC and SNRcc traces coincide as the shorter time window illustrates and the arrival time estimates are identical for both traces.

The SNRcc is an integral characteristic of the similarity between the signal and template, and it is supposed to have an absolute peak near the actual arrival time of the sought signal. Two panels on the right in Figure 10 display the CC and SNRcc traces for the aftershock in Figure 8. There is a clear difference in the peak values timing. The SNRcc peaks 10 counts before the *CC* and a correction of the SNRcc to CC arrival time is needed. We limit the search of the CC-peak to ±1 s around the SNRcc peak. Other search intervals may be also applied. The choice of an optimal search window is case-specific.

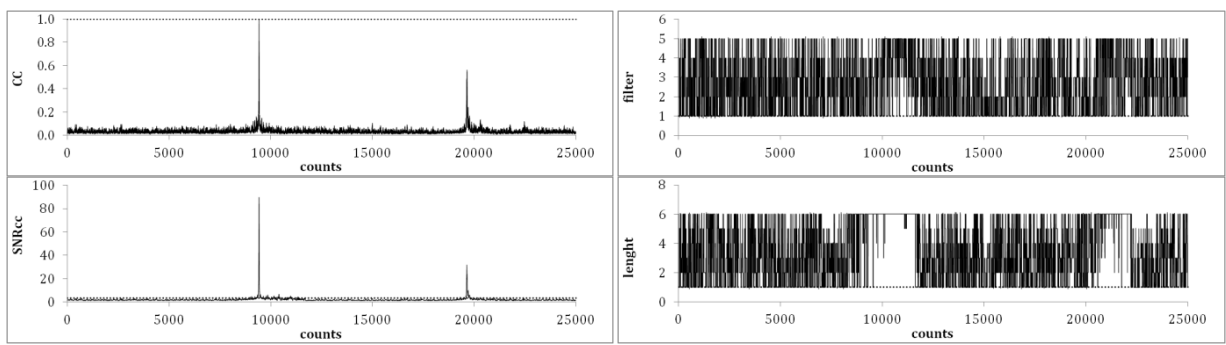

Figure 8. Traces of cross-correlation coefficient, *CC*, the signal-to-noise ratio for the CC-trace, SNRcc, the filter of the largest SNRcc, and the length of CC-window for a given time. The SNRcc for the DPRK6 mainshock (2017246) and the first aftershock is well above the detection threshold (dotted line, SNRcc=3.5). The master event is the DPRK6 and *CC*=1.0 as a result of autocorrelation. SNRcc=90 for *CC*=1.0. For the first aftershock, 510 s or 10200 counts later, *CC*≈0.6 is clearly seen above the noise level but SNRcc=30.

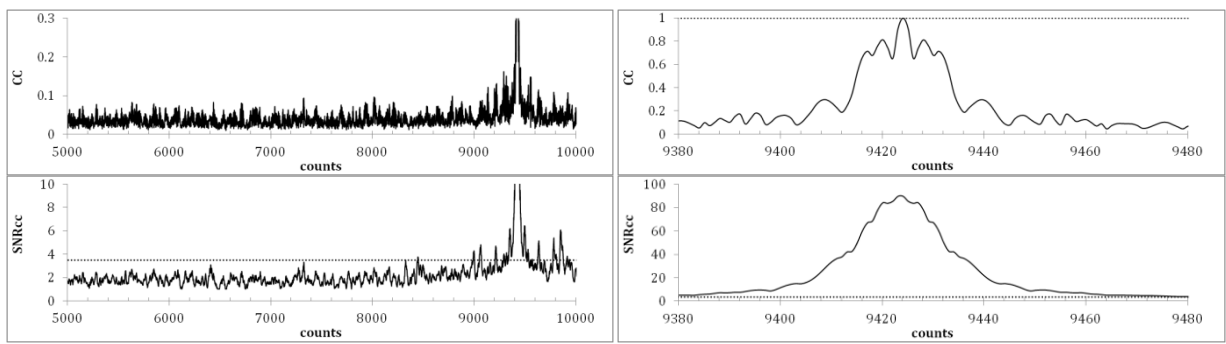

Figure 9. Traces of cross-correlation coefficient, *CC*, the signal-to-noise ratio for the CC-trace, SNRcc, in time intervals of 250 s (left) and 5 s (right) for the mainshock of the DPRK6. The peak SNRcc and CC coincide. The first SNRcc above the threshold of 3.5 is observed ~50 s before the peak.

Figure 11 presents the same four parameters for the same time interval as in Figure 8, but for another date - 2021246, *i.e.* exactly four years after the DPRK6. This was a quiet day without any detected DPRK aftershocks. The trace of absolute CC values does not show any large peaks and the SNRcc has just a few random SNRcc-values slightly above the detection threshold shown by a dotted line (SNRcc=3.5). The winning filters and CWLs are also randomly distributed as expected for a stochastic seismic trace with no coherence with the DPRK templates. For a given array station, the number of random detections depends on the SNRcc threshold and noise properties.

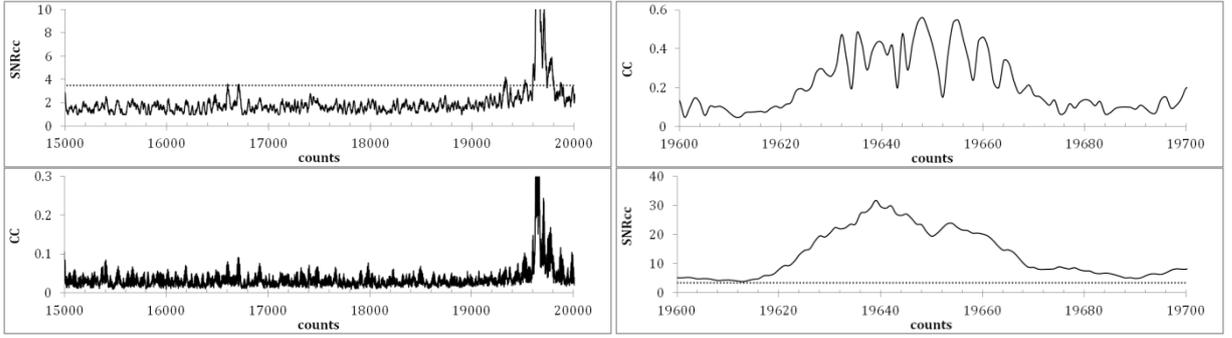

Figure 10. Traces of *CC* and SNRcc, in time intervals of 250 s (left panels) and 5 s (right panel) for the first aftershocks of the DPRK6. The peak SNRcc and CC do not coincide (right panels) and a correction of 10 counts (0.5 s) from the SNRcc peak to the CC-peak is needed for a more accurate arrival time estimate. The first SNRcc above the threshold of 3.5 is observed ~20 s before the peak.

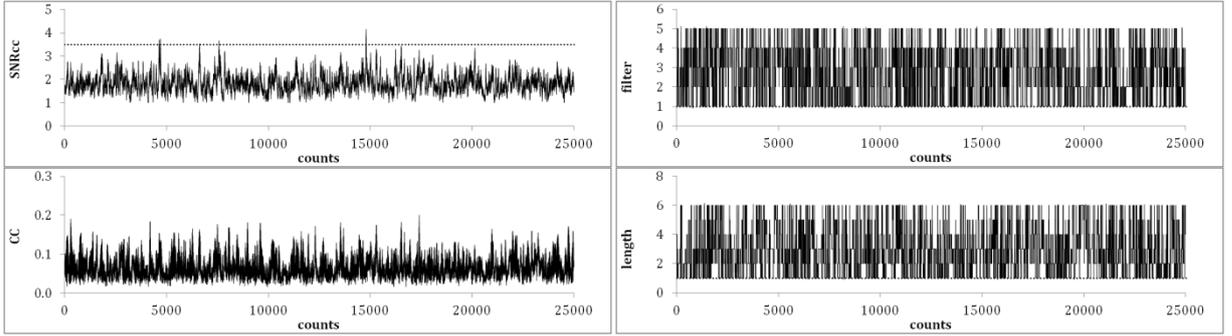

Figure 11. Traces of cross-correlation coefficient, *CC*, the signal-to-noise ratio for the CC-trace, SNRcc, the filter of the largest SNRcc, and the length of CC-window for a given time. The same time intervals as in Figure 8 are selected for the day of September 3, 2021, i.e. four years after the DPRK6. There were no aftershocks detected. The SNRcc above the detection threshold is likely related to noise variations.

The matched filter detector is based on cross-correlation between two signals and represents the optimal linear filter for maximizing the signal-to-noise ratio (SNR) in the presence of additive stochastic noise. Seismic noise might not be fully stochastic, *i.e.* the signals constituting the noise may have characteristics similar to both or one of two signals. (For example, detection of signals from the aftershock zone of the Tohoku earthquake with a matched filter is a challenge.) The presence of coherent signals makes the level of similarity as measured by the cross-correlation coefficient to be less reliable. We use SNRcc to characterize the similarity between two signals: it takes into account both the similarity between signals and suppression of uncorrelated noise as we suggest that the ambient noise has no significant component coherent with any of the found aftershocks. The only exception is the immediate aftershock (8.5 min) of the DPRK6. In this case, the emission from the mainshock and corresponding seismic noise at USRK and KSRS is clearly coherent with the signal from the aftershock. This aftershock is by far the biggest among the DPRK aftershocks.

A detection is declared when SNRcc exceeds the predefined threshold. At this point, the LTA has to be frozen and the SNRcc depends only on the STA throughout the period of 2·CWL starting at the detection point. This procedure is applied to all combinations of the CWLs and filters. The reason to freeze the LTA is simple - one needs to avoid the influence of the signal on the LTA, which has to be the measure related to the ambient noise. Considering the growing SNRcc from the noise level to its peak near the signal arrival time and varying CC-window length, the maximum SNRcc is sought within the length of the CC-window corresponding to the detection point. There are two possibilities: to use the SNRcc trace obtained for the CC-window and filter in the detection point or the maximum SNRcc trace over all available filters and CWLs. For the sake of internal consistency, we use the first option, but also check the result of detection by the second method. After the SNRcc peak is found, the maximum *CC* value is determined within the ±1 s window. This is an important correction of the arrival time because the basic assumption is that the synchronization of signal and template results in the highest *CC* value, which may differ from the peak SNRcc due to the noise influence when the sought signal is weak. The next CC-detection has to be separated from the previous one by the length of the CC-window plus LTA in order to avoid the mutual influence of the sequential signals on the SNRcc. The DPRK aftershocks are likely not so often and close to each other to be affected by the mandatory still period between the subsequent arrivals. For the aftershock sequences of catastrophic earthquakes, such an assumption is likely not appropriate.

There are several parameters estimated for a detection: arrival time, SNRcc, standard SNR, the relative magnitude, the filter and CWL for the detection point. There are optional parameters, which can be calculated for any detection if needed: the FK-analysis is used to estimate slowness and azimuth using the original waveforms, and the same type of FK-analysis can be formally applied to the CC-traces at an array station. When the amplitude of the detected signal is large enough, the FK-analysis can provide accurate estimates of the vector slowness. For weak signals, the standard FK-analysis is less effective, but the FK-analysis applied to the CC-traces is able to provide additional information on the signal quality despite the physical signal is compatible in amplitude to the ambient noise. There is a limit to the efficiency of the CC-FK, however, and for the weakest CC-signals the estimates of azimuth and slowness are also poor. We do not use the FK and CC-FK analysis to screen out poor detections because some of them are valid and can be used to create event hypotheses.

Seismic signal detection is not a simple procedure when applied to the CC-traces. It needs careful tuning. The choice of CWLs, filters, STA, LTA, and other parameters has to be based on extensive statistics gathered in an iterative procedure for each master separately. The optimal set of parameters has to balance the rates of true and false detections, which are declared true and false when these detections are associated with reliable event hypotheses. The meaning of a reliable event hypothesis may change over time and depend on specific tasks. Tuning of IDC routine automatic processing is based on the interactive analysis sorting out true and false event hypotheses after the retiming/parameter re-estimation session and extensive search of potentially missed detections [Coyne *et al.*, 2012]. Such an approach is effective in the long run. The DPRK series of weak aftershocks is a short set of events and detections. Tuning of the CC-detection procedure in this specific case might not provide the optimal setting.

**5. The multi-master method: Local Association, Conflict Resolution, and Event Definition Criteria**

Detection is the first step of the event creation process. A valid event hypothesis has to include several detections with parameters not contradicting the origin time, location, and magnitude of the given event hypothesis. Following standard IDC definitions, we suggest that, when one or more parameters of a detected signal is (are) within the allowed uncertainty limits for a tested event hypothesis, this signal is considered as associated with the event and also is defining if the parameter belongs to the set of defining parameters. The IDC uses three defining parameters: travel time residual, *i.e.* the difference between theoretical and observed arrival times, azimuth, and slowness [Coyne *et al*., 2012]. By design, the WCC method detects only the signals with vector slowness practically equal to that of the master event, and thus, these parameters have to match the association criteria. A crucial difference with the IDC association procedure is that the WCC travel time residual does not use the theoretical travel estimate. The master events provide the extremely accurate, ~0.001 s, empirical travel times. The theoretical travel time uncertainty depends on phase and distance and the value of 1 s is a conservative estimate for the first P-waves at regional distance.

A standard procedure for associating several phases with a unique seismic event is based on the assumption that one arrival of a regular seismic phase at a given station is counted only once. This is a physically and statistically sound assumption. The probability of a hypothesis of an event, *i.e.* the combination of hypocenter and origin time, depends on the probability that several observed phases have close origin times at the same source. For a given station, the origin time is estimated as the difference between the arrival time and the phase-dependent travel time. To ensure a predefined level of reliability for all event hypotheses in the REB, the first P-waves at three or more primary seismic stations have to be associated with any pure seismic (no infrasound or hydro-acoustic arrivals) event. In the standard IDC approach, the quality of each phase associated with an event, and thus, its contribution to the overall probability or the weight of the corresponding event hypothesis, can be estimated only for three parameters: travel time residual, scalar slowness, and azimuth. Other parameters like SNR have no direct value for the EDCs. However, the quality of arrivals is important for location, and thus, has an indirect influence on the event hypotheses.

The IDC's set of event criteria is of limited usefulness for studying the DPRK aftershocks. First of all, we can only use two stations. At the same time, within the first hundreds of kilometers from the test site, there are several mining areas that generate regional $L_g$-waves looking like those observed from the DPRK aftershocks. For signals with small amplitude, it is difficult to distinguish aftershocks from quarry explosions even on such large seismic groups as KSRS. In some cases, only the 5.5 s arrival time difference at KSRS and USRK makes it possible to distinguish aftershocks from quarry blasts or natural seismic events outside the test site.

For recurring events within a few wavelengths of the generated seismic phases (*e.g.*, 2 km for a 4 Hz harmonic of the $P_n$-wave with 8 km/s velocity), one physical signal from one source can be taken into account several times if it is detected by cross-correlation using waveform templates from close but different events. The intuition behind this procedure is that signals from spatially close events are more similar than signals from remote events, especially at seismic arrays, which are seismic antennas. Moreover, a slight change in the template shape between master events can be useful for recognizing whether this change is of physical origin. The number of arrivals detected by a given set of MEs at one station is a good measure of the reliability of the corresponding physical signal, which may be hidden in the microseismic noise. The rate of successful signal detections is somewhat similar to the detection rate of radar pulses as a parameter of radar performance. The difference in the template shapes is somewhat similar to changing pulse width or frequency content. Therefore, we consider the simultaneous association of arrivals detected by several MEs, although they are all associated with one physical signal, as a version of radar operation. The CC-arrivals are similar to the detected reflections from the sought object. To find an object of interest, radar has to sum up many low amplitude reflections and then obtain a reliable signal. To find a weak DPRK aftershock, one has to sum up many low amplitude CC-detections and then obtain a reliable event hypothesis.

For the WCC method applied to the DPRK aftershocks, both seismic antennas, KSRS and USRK, are steered to the test site and we need to define how many detections or associated templates, Nass, is enough to define an event hypothesis as a reliable one. It is also important that the input of each station is significant. The multi-master approach is a version of radar processing with elastic waves instead of electromagnetic and signals from repeating events instead of emitted pulses. In routine seismic processing, seismic antennas work in a passive regime and the multi-master method is rather an active mode version. The best master event from the set we use in this study may probably find weaker aftershocks than the multi-master method, but the latter better balances high resolution and high reliability.

In general, for the success of the multi-master approach, it is important to determine the optimal set of template parameters and determine the detection threshold of the sought signal, for example, as a function of the probability of the hypothesis of a seismic event. For the specific case of the DPRK aftershocks, we have two stations and a limited number of templates that were obtained iteratively from the first aftershock detected on September 11, 2016, using only two templates from the DPRK4. The known start time of the DPRK aftershock sequence makes it possible to determine a quasi-optimal set of detection and association parameters using a predetermined probability of a false event. For example, we allow one false event per month. By varying the detection threshold and the minimum number of associated templates, it is possible to control the rates of detection and event hypotheses.

There are two types of false events: 1) result of a random association of stochastically distributed false detections; 2) false positives created by the detection of irrelevant strong signals from larger events due to side lobes of seismic arrays KSRS and USRK, which are working as antennas for the CC-detector. The first type is usually characterized by low SNRcc values and low relative magnitudes, as reported in [Bobrov *et al*., 2014]. For random CC detections obtained with 57 templates, the probability that a large number of arrivals have close origin times depends on the rate of arrivals and the length of the origin time window. The higher the detection threshold, the lower the false alarm rate of the CC-detector. The larger the required number of arrivals in a given origin time window to construct an event hypothesis, the lower the cumulative probability of such an outcome at a given rate of stochastically distributed detections. We can control the frequency of stochastic false positives and false events.

The second type of false alarm is also characterized by low SNRcc values, but they have much higher relative magnitudes. Relationship (1) shows that, for a given master event, the relative magnitude of a SE is proportional to the logarithm of its RMS amplitude, $|S|$, estimated in the cross-correlation window. The relative magnitude $RM = M_m + dRM$, where $M_m$ is one of the appropriate magnitudes of the master event, *e.g.* $m_b$ or ML. Therefore, the false detections caused by large-amplitude regional and teleseismic P-waves give larger relative magnitudes. The teleseismic P-waves are predominantly associated with impulsive signals, and thus, better correlate with the templates from the DPRK explosions. As a result, we can easily identify such false events and exclude them from the false event rate as irrelevant. For example, we found three false events of type 2 on March 11, 2011, with magnitudes above 3.3. The main shock of the Tohoku earthquake created an enormous number of high-amplitude seismic phases. This is the highest number per day for the conservative CC processing with the target of one false event per month. During the next two days after the main Tohoku shock, we observed 1 false event per day. Then, the Tohoku-related false events were not observed.

The multi-master association procedure uses 29 MEs with 57 templates in total [Kitov *et al.*, 2018; Kitov, 2021]. When cross-correlation and detection are completed and the final list of arrivals is available for further analysis, we reduce all the arrival times to the origin times. For one day of data, we line up all the origin times in the absolute time order. Then the number of arrivals in an eight-second window, which is a predefined length of the association window, AWL, is sequentially counted for each and every reading, from the first to the last minus eight. The length of the association window is an important parameter that should take into account the origin times scattering and the change in travel times to USRK and KSRS for the virtual location grid in the Local Association. The allowed travel time residual is ±3 c and the 8 s AWL accommodates a reasonable level of origin time scattering and travel time corrections. The CC-detection procedure prohibits two arrivals for a given template to be within 30 s. Therefore, only the arrivals obtained by different templates may fall within an 8 s interval.

When the number of arrivals in an 8 s window exceeds a predetermined threshold, the procedure of relative position estimation is triggered as part of the LA process. We move the potential location over the virtual grid and calculate the average origin time for the arrivals within ±3 s of the average. If one or more arrivals are out of the limits, but there are still event hypotheses matching the minimum number criterion, we recalculate the average origin time. To facilitate the spatial grid search we introduce the origin time grid with a 0.1 s step. Therefore, we search through all possible virtual locations and origin times for a given set of arrivals and find the combination of virtual location and average origin time with the highest number of associated templates. This best combination is saved as an event hypothesis and all rival combinations are rejected, i.e. two valid event hypotheses cannot coexist within 8 s interval.

It is possible that two or more combinations of origin time and location in a given association window have the same number of associated templates. Then we chose the one with the lowermost RMS origin time residual. This is a Conflict Resolution procedure similar to that for the standard case adopted in the XSEL processing when one physical arrival allowed per phase per station [Kitov *et al.*, 2019]. It is also possible that event hypotheses in neighboring association windows compete for the same arrival(s). This is a more complicated case because the shared arrival can be associated with one of the rival hypotheses by mistake. Then both hypotheses are true and the problem can be resolved by simple disassociation with one of the two. The rule to select the winning hypotheses is simple - it is more probable that the arrival belongs to the stronger hypothesis - the hypothesis with a larger Nass and lower RMS origin time residual.

The fate of the losing hypothesis depends on its reliability. If it loses one associated arrival and still matches the EDCs, it is retained and two subsequent event hypotheses are saved in the XSEL. Otherwise, it is rejected. The interval between the origin times of such hypotheses is short. The aftershock sequences of the largest earthquakes often have such events close in time, but not necessarily in space. For routine IDC processing, it is a challenge to distinguish between the events consisting of similar arrivals separated by a few seconds. The DPRK aftershocks are close in space and also can be close in time due to the mechanics of the cavity and/or chimney collapse. The length of the cross-correlation window, however, reduces the time resolution to 60-120 s. Due to this time spacing between the allowed origin times, there can be no competing aftershock hypotheses in the neighboring association windows of 8 s (plus 3 s of the origin time tolerance). The multi-master method is also effective for finding small explosions within the DPRK test site, however, which can be conducted within a few seconds. The Conflict Resolution procedure is ready for such challenges.

In some cases, two event hypotheses compete for two or more arrivals. The rule of the winning hypotheses selection is the same, but the decision on the rejection of the losing hypothesis depends not only on the EDCs match but also on the number of lost arrivals. There can be a threshold for the largest number of lost arrivals depending on the statistics of the conflict resolution outcome in comparison with alternative solutions like the REB or LEB. The result of the interactive analysis can be useful for the automatic WCC processing as an independent reference. The most sophisticated case is when several event hypotheses are competing for many arrivals which can also be different for each of the competing pairs of the hypotheses. In this case, we start from the most reliable hypothesis and resolve its conflicts with all other hypotheses. Then the most reliable hypothesis left after the first round is processed the same way, and so on.

In the routine stage of the DPRK data reprocessing, the threshold number of the associated templates is set to 11, which is formally equal to the number of the explosion-related templates in the set. This rule prevents missing small explosion-like events. For a successful event hypothesis, both stations must contribute to the solution. The participation factor should be above 0.3 for $Nass<15$ and 0.25 for $Nass\geq15$. The threshold of 11 is not the decision line between valid and false even hypotheses. We use the hypotheses with $11\leq Nass\leq 20$ as seed events at the high-resolution stage. The true/false event decision line is $Nass=20$, except for the cases of specific interest.

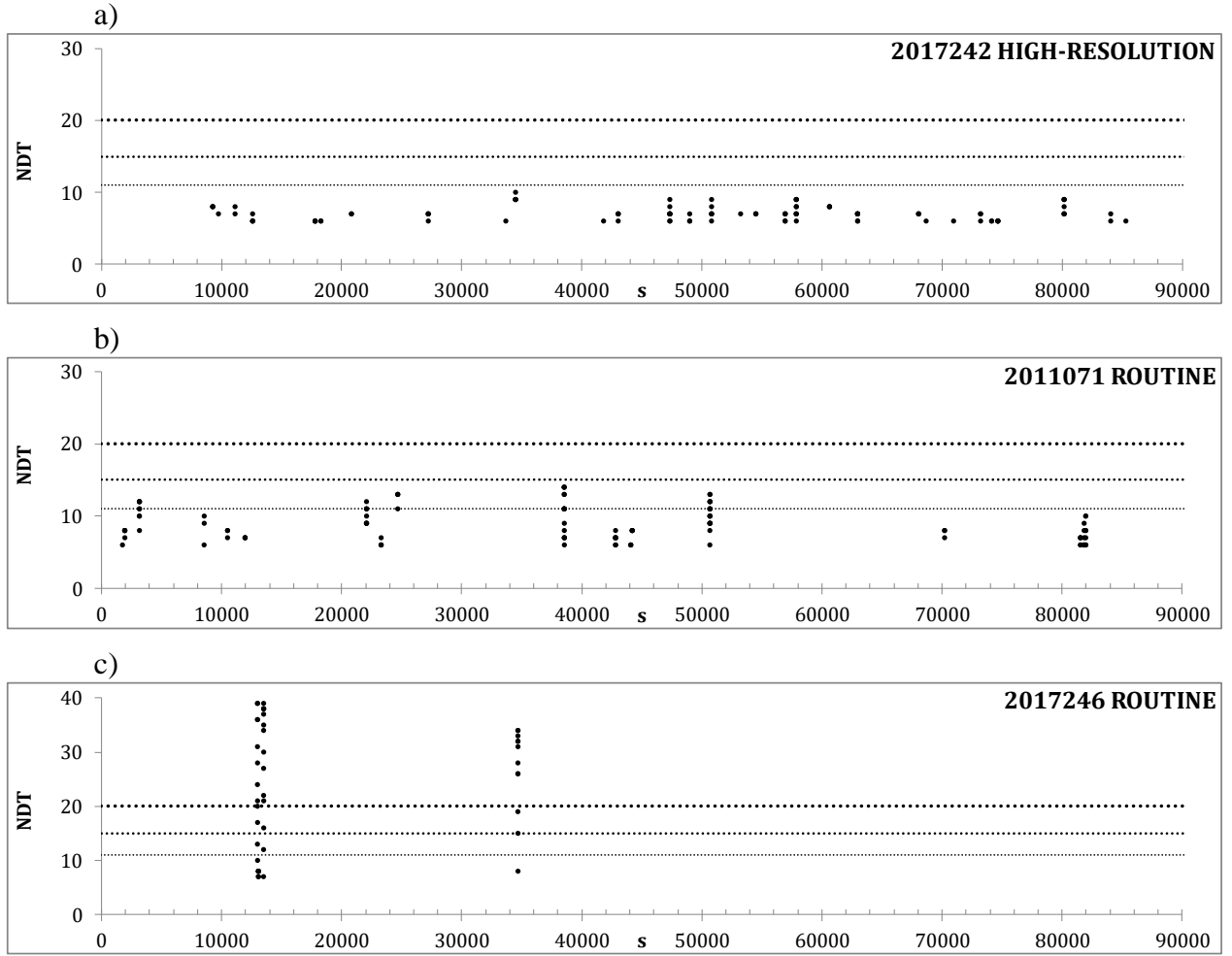

Figure 12. The number of detections, NDT, in the running eight-second-intervals with 1 s step as estimated for 3 different days.

Several specific cases for Local Association and Conflict Resolution processing are illustrated in Figure 12. The number of detections, NDT, made by 57 templates in 9 s intervals starting at subsequent origin times from 0 to 86391 is estimated for 3 different days: a) 2017242; b) 2011071; c) 2017246. We also apply an additional counting rule: each station has an input of 3 or more detections. No aftershocks were found by the multi-master method on August 30, 2017 (2017242). There are no intervals with NDT=11 or more. The intervals with NDT≤11 do not build event hypotheses to be processed with the LA and CR. For 30.08.2017, the result of the high-resolution set of control parameters (as described in the following paragraphs) is shown because there were no intervals with 6 or more detections in the routine WCC processing.

The second day of the aftershock activity related to the Tohoku earthquake (March 12, 2011) is characterized by extremely intensive seismicity at regional distances from the DRPK test site. The number of intervals with 11>NDT≥6 is much larger than that observed on 30.08.2017. There are 4 association intervals with NDT =11 as panel b) shows. Only 3 of them resulted in event hypotheses which were later rejected by their relative magnitudes. One interval did not generate a hypothesis due to other EDCs. Not only the relative location with the virtual grid was used to build these 3 event hypotheses but also the Conflict Resolution. Panel b) shows that there exist neighboring intervals with NDT≥11 and the event hypotheses in these intervals have to compete for the same detections. As a result, only one event hypothesis survives.

The DPRK6 (2017246) had two aftershocks in 8.6 min and 6 h. The number of detections in the association intervals in panel c) gives a clear picture for this day. The association windows with NDT≥6 are observed only around the mainshock and the two aftershocks. The NDT grows to some peak value and then falls back to 0 in several subsequent intervals. The Conflict Resolution has an easy work to find the most reliable event. In the routine WCC processing, no other event hypotheses were created with Nass≥11 because no interval had NDT≥11. This picture demonstrates that even the smallest events can be reliably detected by the multi-master method and the CC-detector has enough resolution and sensitivity to find much weaker events. The absence of weaker events with 35≥Nass≥11 in panel c) indicates the absence such events rather than the limited capability of the method.

After the Local Association process has formed a set of event hypotheses matching the EDCs we check the solutions for possible errors. During this inspection, the events with unreasonably high relative magnitudes are removed from the list as not relevant to the current study. It is also important to check the SNRcc values and remove outliers (unusually high SNRcc) likely related to data quality problems - spikes and the likes. The WCC processing has a special quality control application, but the visual check is not excessive. There is another parameter to check - the total number of detections within the 8 s interval corresponding to a given solution. For weak aftershocks, some arrival times might have low accuracy and occur out of the ±3 s tolerance limits but within the studied time interval. These detections could be added to the given solution if to improve the accuracy of the onset time estimate. The use of the high-resolution set of the WCC control parameters includes this option.

The set of event hypotheses after the visual inspection and rejection of false events is called the cross-correlation standard event list, XSEL, following the naming tradition for IDC products of automatic processing. The XSEL contains the most reliable hypotheses with the event and station parameters. For a given event, the following parameters are estimated: origin time and its standard error, relative coordinates, the relative magnitude with corresponding standard error, zero depth, Nass, etc. Station-related parameters include arrival times, origin times, travel time residuals, SNRcc, *CC*, SNR, d***RM***, etc. It is worth noting that the origin times and arrival times obtained by the WCC method are different from those estimated by standard IDC methods. The weaker the arrival, the larger can be the difference. Such a comparison is possible only for the events with IDC solutions (in the REB, LEB, SEL*), however, which are not many.

In the final, high-resolution, stage of the DPRK-related data reprocessing, the defining parameters for detection and association are chosen to provide consistent estimates of cross-correlation parameters and event hypotheses. All filters and CC-window lengths are used for detection, but only in those hours when arrivals associated with a reliable event hypothesis were detected in the initial processing stage. This allows reducing the required computer resources without losing resolution since the first stage had lower detection and association thresholds. As a result of in-depth analysis, some hypotheses of events obtained in the initial phase turned out to be false and were removed.

The best hypotheses of the first stage can be significantly improved depending on the number of associated arrivals. The local association procedure has been updated to reflect the new distribution of detection parameters, and the minimum Nass value for a valid event is set at 20. This threshold rejects all possible random spurious events. Over the entire observation period, several event hypotheses had 15 to 20 associated phases related to the side lobes of the CC-detector. They were also rejected for their relative magnitudes and domination of the arrivals associated with blast patterns.

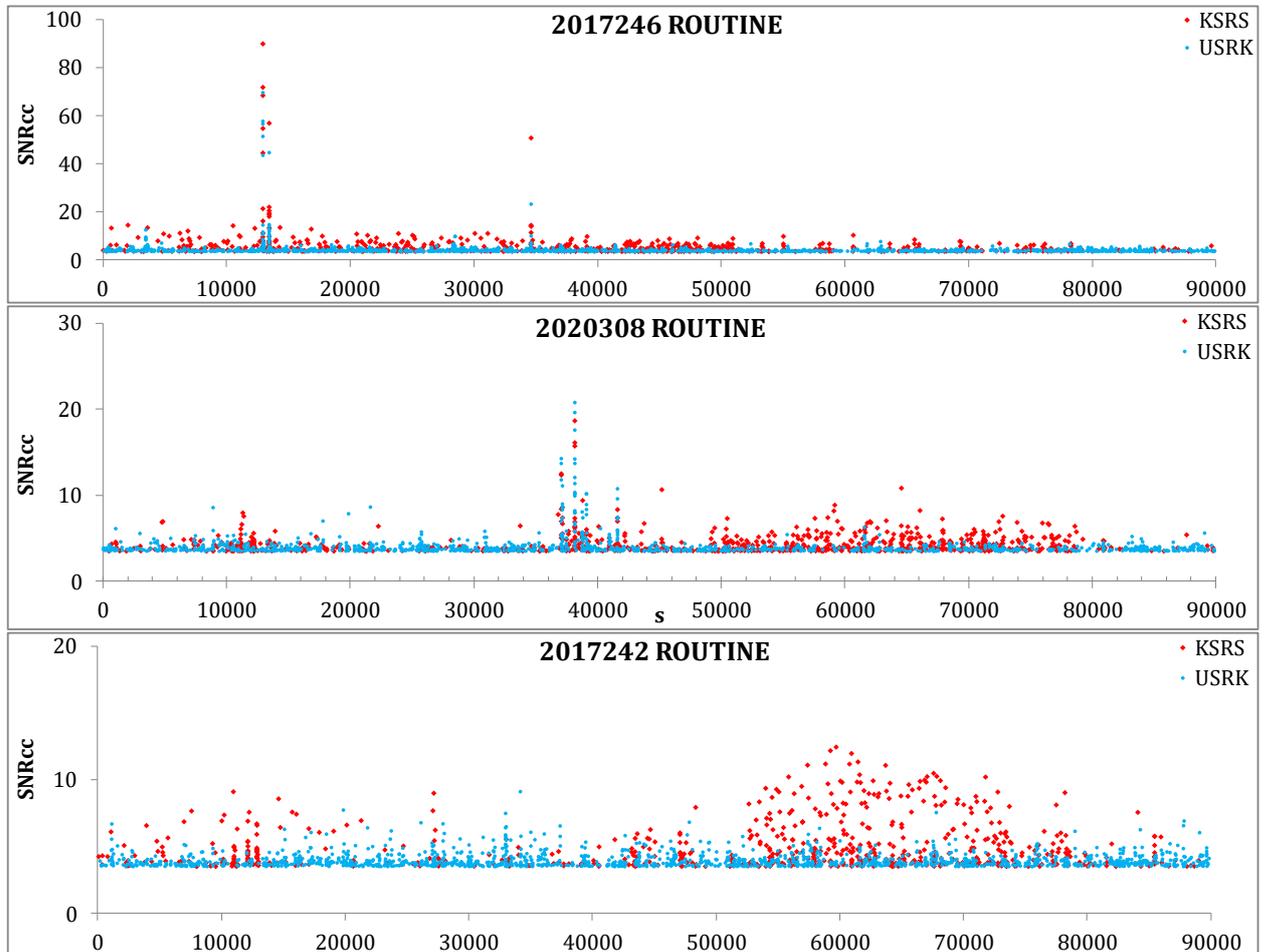

Figure 13. Distribution of SNRcc at stations KSRS and USRK as a function of the elapsed time for a given day.

To visualize the process of Local Association, Figure 13 depicts the origin times for detections, calculated according to relationship (2), at two stations as a function of the elapsed time for a given day. All detections are represented by their SNRcc values. The origin times are obtained with the empirical travel times for the corresponding master events, *i.e.* no virtual location grid is applied. We distinguish detections at KSRS and USRK in order to illustrate the rule that both stations have to provide significant input to a true event. Three different days are presented: September 3, 2017, with the mainshock and two aftershocks clearly seen in the data; November 3, 2020, with 4 aftershocks and more activity between 10 a.m. and 12 a.m.; August 30, 2017, without aftershocks but with many detections having large SNRcc at KSRS in the second half of the day. These large SNRcc detections of unknown origin are not supported by USRK and no event hypotheses were created. Interestingly, a similar SNRcc activity was observed at KSRS on November 3, 2020. The activity begins near midnight (local time UTC+9 h) in both cases. The source is likely natural rather than artificial.

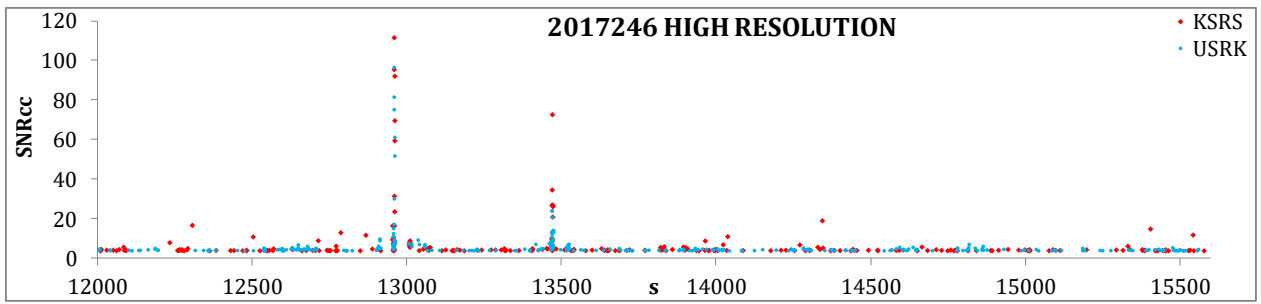

Figure 14. Distribution of SNRcc at stations KSRS and USRK as a function of the elapsed time for one hour with the DPRK6 mainshock and the first aftershock. This is the case of the high-resolution processing (STA=0.5 s, SNRcc threshold 3.6), which is aimed at the increase in the number of associated templates. The SNRcc is increased compared to the solution in Figure 13.

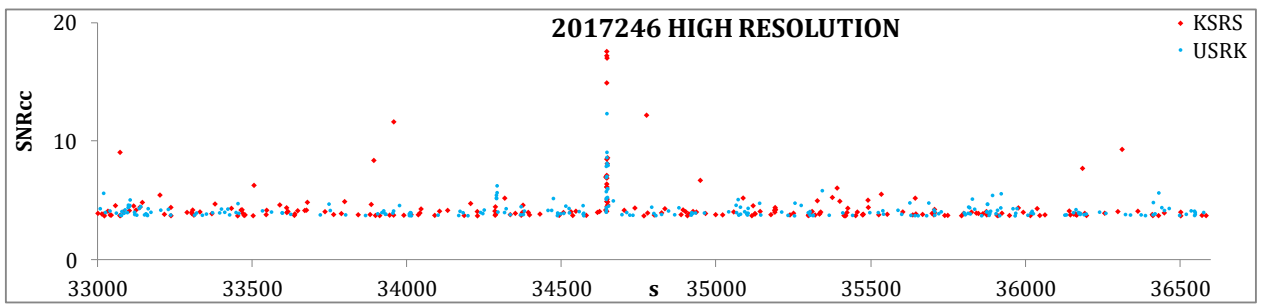

Figure 15. Same as in Figure 14 for the hour with the second aftershock.

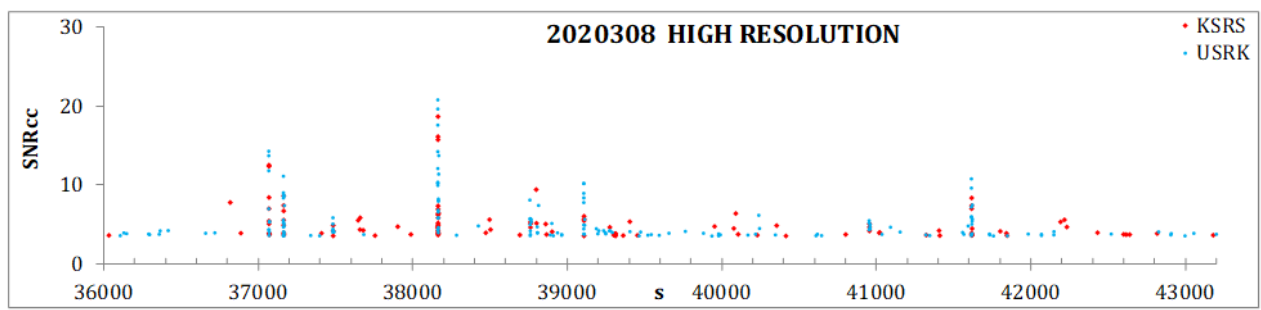

Figure 16. Same as in Figure 14 for two hours of 2020308.

In Figure 14, the distribution of SNRcc at stations KSRS and USRK is presented as a function of the elapsed time for one hour which includes the DPRK6 mainshock and the first aftershock. This is the case of the highest resolution processing (STA=0.5 s and SNRcc threshold 3.6) of one-hour data, which is aimed at the increase in the number of associated templates. The SNRcc values are increased compared to the solution in Figure 13. The number of associated templates is: 38 for the DPRK6 in the routine WCC processing (STA=0.8 s, CWL=120 s, SNRcc threshold 3.5) and 37 in the highest resolution processing; for the aftershock Nass=36 and Nass=37, respectively.

The second aftershock on 03.09.2017, is shown in Figure 15. For this small aftershock, the number of associated templates has increased from 31 to 41. The high-resolution regime is aimed at the improvements in the weakest event hypotheses and might be helpless for the biggest aftershocks and explosions. There are too many detections for different filters and CWLs competing for the same physical signal. The probability of an error in arrival time estimate increases. This effect is addressed in [Kitov, 2021]. Figure 16 presents two hours of 2020308 with the highest aftershock activity. There are 4 events in the XSEL but 2 more events formally matching the Nass≥20 rule could be potentially included in the XSEL if not obtained in the regime with an extremely high rate of false events generation.

The Event Definition Criteria used in this study include several rules.

1) For an associated template, the origin time residual has to be within ±3 s. The origin time tolerance is the same throughout the whole study. No azimuth and slowness tolerances are applied.

2) The minimum Nass depends on the task. For example, Nass=11 in the routine stage and Nass=15 in the high-resolution mode. These thresholds are obtained from the data processing and balance the rates of true/false events.

3) The input of one station must be at least 30% of the total. For the events with 15 and more associated templates, this ratio is reduced to 25%, *i.e.* 4 or more templates have to be associated with a 15-template event. This criterion is aimed at the rejection of the event hypotheses built by one station as a result of some regional source generating $L_g$-waves similar to those generated by the DPRK aftershocks. For example, some mines in the DPRK and China.

4) The difference of origin times averaged separately over all associated templates at KSRS and USRK has to be 5.5s ± 1.5 s. This is an extremely powerful criterion for the Local Association. It limits the area of possible sources to a relatively small ellipse around the DPRK test site. There are no mines around the test site that can match this criterion.

5) All associated signals have to have SNRcc above some threshold dependent on the processing regime. For the routine WCC regime with STA=0.8 s, the SNRcc threshold is 3.5. No threshold is defined for standard SNR since the WCC method can find signals deep in the ambient seismic noise.

6) The relative magnitude of an event hypothesis has to correspond to the strength of the solution as expressed by the number of associated templates. For example, a valid 11-template hypothesis cannot have ***RM***=4. Such an event is most likely related to the side lobes of the CC-detector which may detect high amplitude signals from larger events. There is no strict rule introduced for this criterion because of the absence of appropriate statistics. Corresponding decisions are taken interactively.

7) The final selection of true events after all processing stages is the most strict and only the event hypotheses with Nass=20 or more are retained in the XSEL. This criterion is very conservative and helps to exclude any false event detected before September 9, 2016. Before this date, no DPRK aftershocks were reported in the previous studies. Hence, any new event found before 2016255 has to be extremely reliable.

## 6. Tuning resolution

There are many routine processing parameters controlling the rates of true/false events. In automatic IDC processing, the rate of true and false events is determined by the detection flux and EDCs. The detection rate is mainly controlled by the set of detection thresholds estimated for IMS stations. For array stations, detection thresholds may vary among beams, which are defined by vector slowness and frequency band. For 3-C stations, the threshold depends mainly on filters. The Global Association, GA, procedure links detections with a set of event hypotheses [Coyne *et al*., 2012]. The essence of GA is the possibility of almost any detection to be associated with many events in various regions of the Earth. This creates a large number of possible permutations in the GA process and conflicts between event hypotheses competing for the same detection.

Calculation of CC-traces and detection of physical signals in the cross-correlation domain creates new controlling parameters. The Local Association process is also different. For a given master, the travel time for all detections has to be within a narrow range around the empirical master/station travel time extremely accurately (~0.001 s) measured for the corresponding master event. The performance of the WCC method critically depends on optimal selection of these controlling parameters (*e.g*., parameter definition) and accurate estimation of the ranges where these selected parameters demonstrate the best performance in terms of the rates of true/false detections and event creation. In each specific case, the best performance has to balance the cost of one missed valid event and the cost of false events creation needed to find and add this missed event.

Comprehensive nuclear test monitoring implies a very high cost of one missed event of interest. For a given configuration of the IMS seismic network and available IDC analyst workforce, the task is to find all possible events in the data matching the adopted event definition criteria and confirmed by analysts. For the current study of the DPRK aftershock sequence, we do not use analysts to confirm events since many of the found aftershocks are close or below the ambient noise level. Our approach to event definition criteria is clearly different from those in nuclear test monitoring - we search for robust events in the automatic regime and in many cases ignore potentially valid aftershocks just below the EDCs for the sake of consistent comparison of the events before and after the DPRK tests. The cost of missing a small aftershock after the test is low compared with a false positive event before the tests. In the latter case, one needs to explain the origin of an event demonstrating high cross-correlation with the DPRK aftershocks.

It is worth estimating the probability of a false event created by false detections with a given EDC. The term "false detection" does not mean that the corresponding physical signal does not exist. This term is rather an expression of the fact that physical detections from different physical sources or wrongly identified seismic phases from the same source cannot build a valid seismic event. The term "false event" expresses the fact of wrong association of detections from different sources or wrongly identified seismic phases with one physical event. For the WCC method, we have an extension of this notation by side sensitivity of the CC-detector to strong signals from bigger events. This type of false event is easy to reject since their relative magnitude is much higher than expected from their cross-correlation coefficient.

The estimate of a false event probability is based on a simple assumption of random distribution of false detection over time. Then the probability of one detection to be within an interval with the width equal to the doubled value of the allowed travel time residual is:

$$P_i = 2 \cdot t_{res}/86400$$

where $P_i$ is the probability of a random detection at station *i* to be within $2 \cdot t_{res}$ interval, $t_{res}$ – the allowed origin time residual. This probability is calculated for the daily rate of false detections, where the day length is 86400 s. For N randomly and independently (in IDC seismic processing any detection creates a still period with a duration of 4 to 90 s depending on the parameters of the following detections) distributed detections per day, the probability of at least one from N detections, $P_N$, within a given $2 \cdot t_{res}$ interval can be estimated by a simple relationship:

$$P_N \sim P_i N$$

At the IDC, a valid seismic event has to include detections obtained by three or more seismic stations. Disregarding 19,600 combinations of 3 from 50 (according to the CTBT) IMS primary stations and filtering out of detections by azimuth and slowness before they are associated with a given event, one can estimate the probability of false detection to be within the $2 \cdot t_{res}$ interval as if the number of stations is equal to the minimum number of the associated stations and all detections can be associated. Therefore, the probability of a false event, PFE, is conditional on the number of false detections at each of M stations, $N_m$:

$$PFE \sim \Pi(N_m \cdot P_m) \qquad (7)$$

For $t_{res}=3$ s, $P_m=6/86400 \sim 7 \cdot 10^{-5}$. For 3 stations, the probability of a false event PFE $\sim 3 \cdot 10^{-13}$. For a goal of 1 or fewer false events per day, *i.e.* PFE$<1.0 \cdot 10^{-5}$, it makes the number of allowed false detections ~500 per day per station with the average time interval between subsequent arrivals $\Delta t=180$ s.

In automatic XSEL processing of the prototype pipeline, the virtual location grid in the Local Association process extends the strict time interval $2 \cdot t_{res}$ by allowing different from the master events locations, and thus, master/station travel times. To adjust the processing to this effect and following a more conservative approach than estimated for the random detection distribution, we limit the rate of false detections to the range between 60 and 150 per day. In the XSEL prototype pipeline, we keep the rate of detections in this range. The probability of a false event with a larger number of associated detections drops extremely fast as we multiply PFE by $\sim 10^{-4}$ for each additional station.

For the DPRK aftershock processing, we have only two stations, KSRS and USRK, and 57 waveform templates. The probability of 11 detections made by 11 different templates to be within the same 6 s interval is straightforward $(7 \cdot 10^{-5})^{11} \sim 2 \cdot 10^{-46}$. For 57 different templates, we have approximately $(57!)/(46! \cdot 11!) \sim 2 \cdot 10^{11}$ combinations of 11 different templates. Therefore, one false event with 11 associated detections obtained by different templates has a probability of PFE$\sim 4 \cdot 10^{-35}$ for a random and independent distribution of false detections within 86400 s. When several detections are made by one template, the probability increases following relationship (7). For the sake of simplicity, we suggest that all templates create an equal number of detections per day, N. Then, relationship (7) allows to find the number of false detections per day per template which create, on average, only one false event per day:

$$10^{-5} \sim 4 \cdot N^{11} \cdot 10^{-35}$$

or ~400 detections per day. The average interval between subsequent detections for one template has to be 220 s. There is a significant influence of the location grid on the PFE. The grid allows additional origin time scattering by 1 s to 2 s relative to the master event location. When we use 8 s origin time interval instead of 6 s, the probability $P_i$ is $10^{-4}$ and an event with 11 associated detections within the same 8 s interval is $10^{-44} \cdot 2 \cdot 10^{11} = 2 \cdot 10^{-33}$. Then the PFE of $10^{-5}$ corresponds to approximately 250 detections per day or $\Delta t=350$ s. This is a crude estimate which ignores the minimum distance between detections of the length of CWL for a given detection (start of detection with a given CWL suppresses all other detection lines) and mandatory shift of 30 s to the start of the next detection search. This makes the random detection rate lower and the distance between detections longer on average. There is an upper limit of detections of 2880, *i.e.* each 30 s. For the CWL of 120 s, there are only 720 detections per day.

Another important factor affecting the CC-detection rate is related to the rule that both stations have to add to the event. Therefore, the number of combinations is lower than for a simple set of equivalent members. Since the number of valid combinations of detections drops one can expect a lower rate of false event creation than for the case of all detections are equivalent. It is also important to stress that the similarity between templates at the same station is extremely high and their correlation with the same signal in the studied data might be not a fully random process. For a signal from a DPRK aftershock, one can expect that many templates will detect it. These detections are not randomly distributed anymore. The same effect may cause the side sensitivity of the CC-detector to strong signals from big seismic events.

There are other physical properties of seismic wave-field and processing parameters that may potentially affect the randomness and inter-dependence of the detection set. The basic estimated rate of detections generating one false event per day is likely between 200 to 400 per day per template. This is equivalent to $\Delta t$ of 220 s to 450 s. One can empirically find the corresponding set of defining parameters in the cross-correlation processing using real data and the set of master events/templates. The rate of detection is most dependent on the Short-Term-Average and the detection threshold. Variation of these parameters within wide ranges allows finding the most efficient set, which guarantees a low false event rate and low missed event rate. The latter is difficult to obtain for natural seismic events which have no observable start point in history. Fortunately, the aftershocks of underground explosions have a definite start point and the study of the pre-event period allows to estimate the false event rate without any missed events, which are physically absent.

We vary the principal parameters in WCC processing and estimate their working ranges demonstrating the best result for various purposes. The calculation of the cross-correlation coefficient, *CC*, is the first step. By definition, one needs a template to calculate the *CC*. We have 57 templates from 29 masters events at two stations USRK and KSRS. All these templates were obtained from the events found before April 23, 2018. The aftershocks after this date, even if they are big enough, are not used as masters. The biggest from these templates belong to the master events found in routine IDC processing (see Table 2). They were used as templates to find the mid-size and the weakest aftershocks in the set of 29 masters. Several examples of waveform templates are presented in Figures 4 to 7. The length of templates is 205 s including 5 s before the arrival time at the central elements of IMS arrays USRK and KSRS. The shape of templates from six explosions and 23 aftershocks are quite different. There are also some differences between templates likely related to DPRK5 and DPRK6 [Kitov, 2021]. Therefore, to achieve the best performance, cross-correlation coefficients have to be calculated with different correlation window lengths, CWL, in order to include various parts of the templates corresponding to the whole variety of sources. Filtering of the original waveforms also improves the estimates of cross-correlation coefficients, and thus, signal detection.

In the routine WCC processing, we use a constrained set of parameters that may have different impact on the overall performances. In the high-resolution processing, a broader set of defining parameters is used in order to improve the aftershock solutions. For the optimization of the WCC processing, it is helpful to understand the effect of each parameter separately. For a discrete set of values for a given parameter, it is important to evaluate of effect of any single value. There are many aftershocks of the DPRK tests found in the previous studies, including those obtained by the multi-master method. The DPRK explosions are also well detected by various WCC-based methods. The multi-master method is one of them. It provides the number of associated templates as the principal parameter defining the reliability of any detected event. It is instructive to reprocess these well-constrained events and to evaluate the performance of various combinations of the control parameters.

To begin with, we processed the DPRK6 and two aftershocks on September 3, 2017. The *CC* is calculated for various lengths of correlation window, CWL, for five filters separately, and, for all the five together. Table 5 lists the results of processing. For all cases, the STA=0.8 s, LTA=120 s, and the origin time tolerance $t_{res}$=3 s. The CWL varies from 20 s to 180 s. For each filter, the number of associated templates is presented. In some cases, no event hypothesis with Nass=11 or more is created and "0" value is assigned to these cases. For the mainshock (event 2017246_0), the largest number of associated templates is observed for CWL=180 s and filter #3 - 38. When all five filters are used together the best hypothesis with Nass=41 is created for CWL=20 s. It is a reasonable result considering the impulsive character of the mainshock signal. For the biggest aftershock (event 2017246_1), there are several best CWL/filter combinations with Nass=32. For all five filters, the best CWL=20 s, *i.e.* the same as for the mainshock. These two events seem to be very similar in shape of signals observed at IMS stations. The second aftershock (2017246_2) is different. It is much smaller in amplitude and seems to be less similar to the other two. The best combination is CWL=180 s and filter #3 with Nass=26. For all filters, the same CWL=180 s is the most prolific with Nass=33 templates. The second aftershock is more similar to the signals related to the DPRK5 [Kitov, 2021].

Table 5. Nass as a function of the correlation window length, CWL, from 20 s to 120 s for the main shock and two closest in time aftershocks (3:39:30 and 9:31:28 UTC) of the DPRK6 (September 3, 2017). The STA=0.8 s, LTA=120 s, $t_{res}$=3s. The number of associated templates has a peak for filter #3 (2-4 Hz) for all three events and CWL=180s, which is extremely long for routine processing.

|        | Filter |    |    |    |    |        |
|--------|--------|----|----|----|----|--------|
|        | 1      | 2  | 3  | 4  | 5  | 1 to 5 |
| CWL, s | Main shock DPRK6, 2017246_0 | | | | | |
| 20     | 22     | 27 | 32 | 31 | 31 | 41     |
| 60     | 25     | 33 | 35 | 29 | 30 | 40     |
| 120    | 24     | 32 | 35 | 33 | 32 | 39     |
| 180    | 25     | 36 | 38 | 35 | 34 | 40     |
|        | Biggest aftershock, 2017246_1 | | | | | |
| 20     | 18     | 25 | 26 | 32 | 27 | 37     |
| 60     | 17     | 26 | 27 | 31 | 23 | 34     |
| 120    | 26     | 28 | 32 | 32 | 26 | 36     |
| 180    | 25     | 29 | 32 | 32 | 30 | 34     |
|        | DPRK6 aftershock, 2017246_2 | | | | | |
| 20     | 0      | 0  | 12 | 0  | 0  | 19     |
| 60     | 0      | 11 | 13 | 0  | 0  | 23     |
| 120    | 15     | 23 | 24 | 21 | 18 | 32     |
| 180    | 14     | 21 | 26 | 21 | 18 | 33     |

November 3, 2020, (2020308) was the most prolific day with 4 reliable aftershocks and many aftershock hypotheses with Nass from 11 to 19 that were rejected. We have found that all four events belong to the DPRK5 aftershock sequence [Kitov, 2021]. Table 6 summarizes the estimates for 2 hours, which include all 4 events. The main difference between the mainshock and its immediate aftershocks consists in poor performance of short CWLs: no events are found. The largest number of associated templates is observed for filters #3 and #4 and CWL=120 s. The usage of all five filters creates much more reliable event hypotheses. For WCC routine processing we decided to use filters #2 through #4 and CWL= 120 s. This set balances the calculation time and resolution of the multi-master method.

Table 6. The number of templates detecting (associated with) a given event as a function of the correlation window length, CWL, from 20 s to 180 s for four aftershocks detected on November 3, 2020 (day 2020308). The STA=0.8 s, LTA=120 s. The number of associated templates has a peak for filters 3 (2 Hz to 4 Hz) 4 (3 Hz to 6 Hz) and CWL=120 s with the only exclusion for aftershock 2020308_3, which is the biggest for this day.

|  | Filter | | | | | |
|---|---|---|---|---|---|---|
|  | 1 | 2 | 3 | 4 | 5 | 1 to 5 |
| CWL, s | 2020308_1 | | | | | |
| 20 | 0 | 0 | 0 | 0 | 0 | 0 |
| 60 | 0 | 0 | 0 | 0 | 0 | 0 |
| 120 | 0 | 0 | 0 | 16 | 15 | 21 |
| 180 | 0 | 0 | 0 | 14 | 14 | 17 |
|  | 2020308_2 | | | | | |
| 20 | 0 | 0 | 0 | 0 | 0 | 0 |
| 60 | 0 | 0 | 0 | 0 | 0 | 0 |
| 120 | 0 | 0 | 14 | 20 | 19 | 27 |
| 180 | 0 | 0 | 0 | 19 | 17 | 22 |
|  | 2020308_3 | | | | | |
| 20 | 0 | 0 | 0 | 0 | 0 | 0 |
| 60 | 0 | 0 | 0 | 0 | 0 | 0 |
| 120 | 0 | 23 | 33 | 34 | 30 | 39 |
| 180 | 0 | 23 | 34 | 33 | 27 | 40 |
|  | 2020308_4 | | | | | |
| 20 | 0 | 0 | 0 | 0 | 0 | 0 |
| 60 | 0 | 0 | 0 | 0 | 0 | 0 |
| 120 | 0 | 0 | 12 | 18 | 15 | 21 |
| 180 | 0 | 0 | 13 | 16 | 0 | 19 |

The next parameter to test is the LTA. Using the 2020308 aftershocks, all five filters together, and the same set of other parameters as in Tables 5 and 6, we vary LTA from 20 s to 180 s for two STA values: 0.5 s and 0.8 s. For KSRS, these STA lengths are 10 and 16 readings of the CC-trace; for USRK - 20 and 32, respectively. The rise times observed in the CC-traces presented in Figures 9 and 10 indicate that such STA values are reasonable. The effect of STA on detection rate is much higher and we consider it later. Here, we assess the joint effect of LTA and STA on two target values: 1) the number of templates associated with reliable events; 2) the number of false events created for the whole processed period. The number of false events is calculated for the events with Nass between 11 and 19. Table 7 lists the principal results. In terms of the associated templates, three shorter LTA windows between 20 s and 60 s outperform the longer ones for STA=0.5 s. For STA=0.8 s, the most effective LTA window is 120 s.

There is another target value, however, which demonstrates a well-known tendency of the shorter STA windows to generate false events. For STA=0.8 s, the number of false event hypotheses is lower for all LTAs, with the minimum of 1 per day reached for the LTA windows of 20 s and 40 s. For LTA=120 s, the number of false events is 2. For the routine WCC processing, we use the LTA window of 120 s as balancing the high number of associated templates and the low number of false event hypotheses. In any case, the effect of LTA on the true/false detection rate is lower than that of the STA, CWL, and detection threshold. We do not vary the LTA in the routine and high-resolution WCC processing.

The effect of STA and detection threshold on the number of detections, and thus, the rates of true and false events is the most prominent. The number of detections grows exponentially with decreasing detection threshold as Figure 17 shows for the STA values in the range from 0.5 s to 0.9 s. The detection statistics aggregate all 57 templates and are obtained for the extended set of control parameters: filters #1 to #5, CWL from 20 s to 120 s with a 20 s increment. In the left panel, the case of the mainshock is presented. The time interval is from 3:00 a.m. to 10:00 a.m. The right panel presents the detection statistics for a full day without aftershocks. August 30, 2017, was selected, *i.e.* four days before the DPRK6. For a given STA, the number of detections increases exponentially with decreasing threshold; notice the logarithmic scale. For the shortest STA window of 0.5 s, the satiation effect starts from threshold 3.3. The satiation effect is likely caused by the shortest allowed time spacing of subsequent detections of 30 s plus the length of winning CWL.

Table 7. The number of templates detecting a given event as a function of the LTA from 20 s to 120 s for four events observed on 2020308. The STA=0.5 s and 0.8 s, CWL=120 s, $t_{res}$=3 s. The number of false events is calculated for the events with Nass between 11 and 19.

|  | LTA, s | 20 | 40 | 60 | 80 | 100 | 120 |
|---|---|---|---|---|---|---|---|
| STA=0.5s | 2020308_1 | 28 | 33 | 34 | 32 | 31 | 32 |
| STA=0.5s | 2020308_2 | 29 | 33 | 32 | 31 | 30 | 30 |
| STA=0.5s | 2020308_3 | 42 | 40 | 40 | 41 | 40 | 40 |
| STA=0.5s | 2020308_4 | 27 | 29 | 30 | 28 | 30 | 30 |
|  |  |  |  |  |  |  |  |
| STA=0.8s | 2020308_1 | 23 | 25 | 22 | 23 | 23 | 23 |
| STA=0.8s | 2020308_2 | 24 | 25 | 25 | 26 | 26 | 26 |
| STA=0.8s | 2020308_3 | 38 | 40 | 40 | 40 | 40 | 40 |
| STA=0.8s | 2020308_4 | 20 | 19 | 19 | 20 | 22 | 24 |
| # FALSE EVENTS PER DAY | | | | | | | |
| STA=0.5s |  | 4 | 4 | 4 | 4 | 5 | 5 |
| STA=0.8s |  | 1 | 1 | 3 | 3 | 3 | 2 |

The effect of STA on the detection statistics is illustrated in Figure 18. Two different days with (2020308) and without (2017247) aftershocks are presented. The detection threshold varies between 3.2 and 3.9, as in Figure 17. For shorter STA values, a quasi-exponential dependence is observed in both cases. There is a kink in the curves around STA=0.7 s, which becomes more and more prominent with an increasing detection threshold. This effect indicates that there are signals, true or false in terms of the adopted EDC, with SNRcc above the 3.7 to 3.9 threshold. These detections are related to real physical signals and their quantity does not depend on STA. In turn, the STA is important for the generation of random detections related to the ambient noise. The kink separates the STA range with random or noise detections and real detections, which share increases with the detection threshold. For the days with the DRPK aftershocks, these real detections can build real aftershocks. For the days without aftershocks, the real detections may (due to the side lobes) build false events, like those observed on March 11, 2011, or fail to build any events.

The detection threshold may be used to draw a statistical boundary between the random and real detections: for lower thresholds, the number of random detections is large for any STA and the largest threshold suppresses the generation of random detections retaining the real ones. The kink in the STA curves is an important observation that might be useful for other WCC applications. The results presented in Figures 17 and 18 are case-specific: we use two stations at regional distances and CWL from 20 s to 120 s appropriate for regional seismic phases. For different CWL and seismic phases, one may find that different STA and detection thresholds are needed to balance the rates of noise and real detections. The kink in the STA curves is a good indicator of the transition to real detections. The tasks formulated in Introduction suggest different WCC regimes have to be used to match specific targets. For routine processing, one has to suppress the rate of noise detections. To associate the highest number of templates with a given event hypothesis, one can ignore the generation of false detection/events and use the set of control parameters with the highest resolution, which should not affect the rate of valid detections, however.

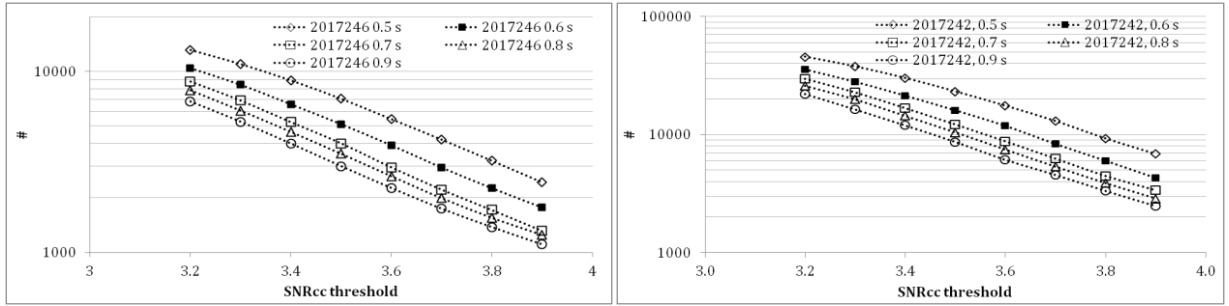

Figure 17. The number of detections as a function of detection threshold for STA from 0.5 s to 0.9 s. Left panel: September 3, 2017, time interval 3:00 a.m. to 10:00 a.m. Right panel: August 30, 2017, full day. For a given STA, the number of detections increases exponentially with decreasing threshold. For the shortest STA window of 0.5 s, the satiation effect starts from threshold 3.3 which is caused by a limit of the shortest time spacing of subsequent detections of 30 s plus the length of winning CWL.

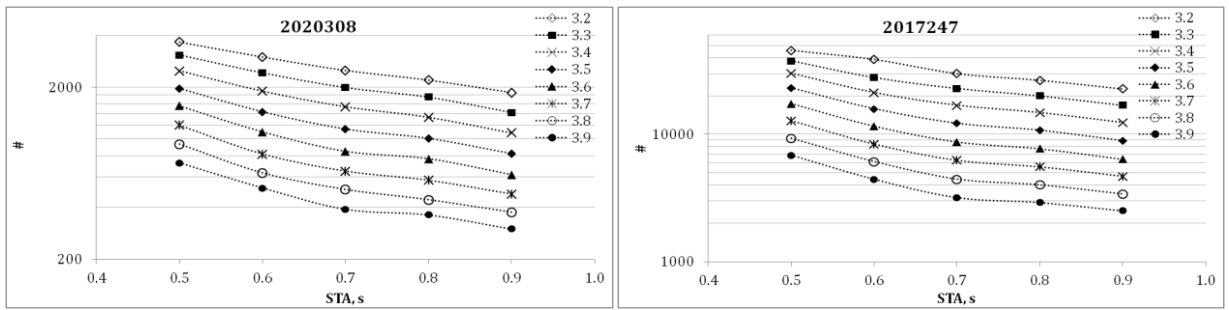

Figure 18. The number of detections as a function of STA for the threshold from 3.2 to 3.9. Left panel: November 3, 2020, time interval 10:00 a.m. to 12:00 a.m. Right panel: September 4, 2017, full day. For a given STA, the number of detections increases exponentially with decreasing threshold. For the shortest STA window of 0.5 s, satiation effect starts from threshold 3.3 which is caused by a limit of the shortest time spacing of subsequent detections of 30 s plus the length of CWL.

An example of high-resolution processing is presented in Table 8. Two days with many known events are selected: 2017246 and 2020308. As discussed above, only the hours with known events are processed. Two parameters are calculated: the number of false event hypotheses created for two thresholds - 15 and 20 associated templates, and the number of new valid events created. The range of STA is 0.5 s to 0.9 s, and the range of the SNRcc threshold is from 3.2 to 3.9. For the STA=0.5 s and SNRcc threshold 3.2, the number of false events with Nass<20 is 22, and 21 of them have Nass<15. This is an extremely large number considering that it was obtained during 7 hours between 3 a.m. and 10 a.m. The same control parameters give 6 false events generated between 10 a.m. and 12 a.m. on November 3, 2020. By scaling this value to the 7-hour period one obtains 21 false events with Nass<20, i.e. practically the same rate as on 2017246. There are 4 false events with Nass<15. This makes 14 false events when scaled to 02.09.2017 or 2/3 of that observed on 03.09.2017. The deficit of events with low Nass on 03.11.2020 is likely caused by the absence of seismic noise generated by the mainshock.

Table 8. The number of false and valid events created by cross-correlation as a function of STA and SNRcc threshold for two segments with the highest aftershock activity as observed between 3 am and 10 am on September 3, 2017 (7 hours) and between 10 am and 12 am on November 3, 2020 (2 hours). False events have two categories: between 11 and 14 associated templates, considered as fully random or related to side sensitivity) and from 15 to 19 associated templates, considered as potentially valid events for further study. The events with 20 and more associated templates are potentially valid event hypotheses.

| 2017246 (3-10) | STA, s | | | | | 2020308 (10-12) | STA,s | | | | |
|---|---|---|---|---|---|---|---|---|---|---|---|
| | False, Nass <20 | | | | | | False, Nass <20 | | | | |
| | 0.5 | 0.6 | 0.7 | 0.8 | 0.9 | | 0.5 | 0.6 | 0.7 | 0.8 | 0.9 |
| 3.2 | 22 | 7 | 5 | 4 | 0 | 3.2 | 6 | 4 | 4 | 4 | 2 |
| 3.3 | 10 | 3 | 3 | 1 | 0 | 3.3 | 3 | 4 | 4 | 2 | 2 |
| 3.4 | 5 | 0 | 0 | 0 | 0 | 3.4 | 4 | 4 | 5 | 1 | 3 |
| 3.5 | 2 | 0 | **0** | 0 | 0 | 3.5 | 4 | 3 | 3 | 2 | 2 |
| 3.6 | 1 | 0 | 0 | 0 | 0 | 3.6 | 4 | 3 | 4 | 3 | 1 |
| 3.7 | 0 | 0 | 0 | 0 | 0 | 3.7 | 4 | 4 | 3 | 1 | 1 |
| 3.8 | 0 | 0 | 0 | 0 | 0 | 3.8 | 4 | 3 | 3 | 2 | 0 |
| 3.9 | 0 | 0 | 0 | 0 | 0 | 3.9 | 4 | 4 | 4 | 1 | **0** |
| | False, Nass <15 | | | | | | False, Nass <15 | | | | |
| 3.2 | 21 | 7 | 5 | 4 | 0 | 3.2 | 4 | 2 | 2 | 2 | 1 |
| 3.3 | 9 | 3 | 3 | 1 | 0 | 3.3 | 1 | 1 | 1 | 1 | 1 |
| 3.4 | 4 | 0 | 0 | 0 | 0 | 3.4 | 1 | 2 | 3 | 0 | 3 |
| 3.5 | 2 | 0 | **0** | 0 | 0 | 3.5 | 2 | 1 | 2 | 1 | 2 |
| 3.6 | 1 | 0 | 0 | 0 | 0 | 3.6 | 1 | 2 | 3 | 0 | 1 |
| 3.7 | 0 | 0 | 0 | 0 | 0 | 3.7 | 1 | 3 | 3 | 1 | 1 |
| 3.8 | 0 | 0 | 0 | 0 | 0 | 3.8 | 2 | 2 | 3 | 2 | 0 |
| 3.9 | 0 | 0 | 0 | 0 | 0 | 3.9 | 1 | 1 | 4 | 1 | **0** |
| | Valid, Nass >19 | | | | | | Valid, Nass>19 | | | | |
| 3.2 | 0 | 0 | 0 | 0 | 0 | 3.2 | 2 | 2 | *1* | 0 | 0 |
| 3.3 | 0 | 0 | 0 | 0 | 0 | 3.3 | 2 | 1 | 0 | 0 | 0 |
| 3.4 | 0 | 0 | 0 | 0 | 0 | 3.4 | 1 | 1 | 0 | 0 | 0 |
| 3.5 | 0 | 0 | 0 | 0 | 0 | 3.5 | 1 | 1 | 0 | 0 | 0 |
| 3.6 | 0 | 0 | 0 | 0 | 0 | 3.6 | 1 | 1 | 0 | 0 | 0 |
| 3.7 | 0 | 0 | 0 | 0 | 0 | 3.7 | 1 | 0 | 0 | **0** | 0 |
| 3.8 | 0 | 0 | 0 | 0 | 0 | 3.8 | 0 | 0 | 0 | 0 | 0 |
| 3.9 | 0 | 0 | 0 | 0 | 0 | 3.9 | 0 | 0 | 0 | 0 | 0 |

The signals generated by the aftershocks that occurred on November 3, 2020, do not interfere and suppress each other. No valid aftershocks were added on 03.09.2017 and 2 new events with Nass≥20 were built on 2020308 in addition to 4 events in the XSEL. These events are likely physically valid, considering the pattern of detections in Figure 16. Figure 19 illustrates the extreme case with STA=0.5 s and SNRcc threshold of 3.2 for the same time intervals as in Table 8. The number of detections, NDT, is calculated in the running eight-second-interval with 1 s step. There are no additional NDT≥20 for 03.09.2017 except those in the XSEL. Two additional cases with NDT≥20 are clearly seen 03.11.2020. These potential event hypotheses disappear with increasing detection threshold and STA.

The generation of dozens of false events per day is a negative result, however, and one needs to avoid such an outcome when processing a dozen years. An optimal set of control parameters has to be selected from Table 8. The rate of false events has to be zero for one day. Moreover, we prefer a conservative approach: the STA and detection threshold have to be one-tenth higher than the optimal ones generating zero false events per day. For 03.09.2017, the set STA=0.8 s and SNRcc=3.5 is a conservative estimate highlighted in bold. For 2020308, the highest real aftershock activity made STA=0.9 s and SNRcc=3.9 almost optimal. Larger values out of the studied range are needed for both control parameters. For both days, the optimal set has no valid events added.

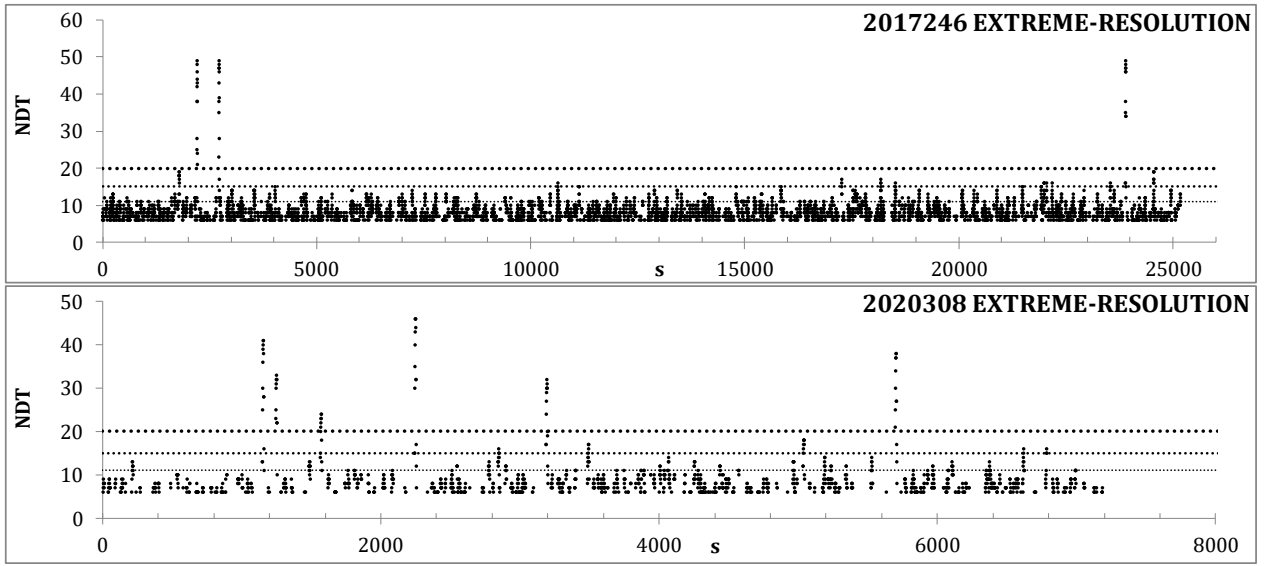

Figure 19. The number of detections, NDT, in the running eight-second-intervals with 1 s step as estimated for 2 different days: 2017246 from 3 a.m. to 10 a.m., 2020308 from 10 a.m. to 12 a.m. The extreme set of the control parameters is used: STA=0.5 s and SNRcc threshold is 3.2 in both cases.

The days before the DPRK6 should not be affected by real aftershocks, and the results of WCC processing are not corrupted. Table 9 presents the results of processing for August 31 and September 1, 2017. The number of false events for the most prolific set of control parameters STA=0.5 s and SNRcc threshold of 3.2 is 45 and 40, respectively. When scaled to 7 hours, the number of false events is 14 and 11. These estimates are approximately two times lower than for the days with aftershocks. The difference is likely related to the fact that there are many signals on 03.09.2017 and 03.11.2020 which are coherent with the aftershock templates and the rate of detections is the same, but more additional poor event hypotheses are created. The list of false event hypotheses for 31.08.2017 and 01.09.2017 is dominated by poor events, with Nass<15. No events with Nass≥20 are created. For both days, the optimal set of control parameters corresponds to STA=0.8 s and the SNRcc threshold of 3.5.

For a random and independent distribution of (false) detections, there should be no true events and the number of false events, on average, has to be less than 1 per day. Instead of using the number of detections per day, we estimate the average time spacing between the subsequent detections, $\Delta t$, for one master event, as discussed at the beginning of this section. This makes it easier to compare daily estimates with the estimates obtained in the narrow time intervals of the highest aftershock activity. Table 10 lists the $\Delta t$ estimates for such active intervals on 03.09.2017 and 03.11.2020. When compared with Table 8, the results in Table 10 for 03.09.2017 indicate that $\Delta t$ of 550 s corresponds to the conservative set of parameters STA=0.8 s and SNRcc threshold 3.5. For 03.11.2020, the optimal set is extremely conservative with $\Delta t>1400$ s. This is the effect of real activity observed in the studied time interval. Effectively, this conservative set of parameters suppresses the creation of real event hypotheses.

Table 9. The number of false events created by cross-correlation as a function of STA and SNRcc threshold for two full days (25 hours - the signals for events around midnight can be detected the next day) without aftershocks found in this study. False events have two categories: between 11 and 14 associated templates, considered as fully random or related to side sensitivity, and from 15 to 19 associated templates, considered as potentially valid events for further study.

| 2017244 | STA, s | | | | | 2017243 | STA, s | | | | |
|---|---|---|---|---|---|---|---|---|---|---|---|
| | False, Nass <20 | | | | | | False, Nass <20 | | | | |
| SNRcc threshold | 0.5 | 0.6 | 0.7 | 0.8 | 0.9 | SNRcc threshold | 0.5 | 0.6 | 0.7 | 0.8 | 0.9 |
| 3.2 | 40 | 17 | 7 | 5 | 2 | 3.2 | 45 | 18 | 10 | 5 | 2 |
| 3.3 | 20 | 8 | 3 | 2 | 0 | 3.3 | 24 | 10 | 3 | 1 | 0 |
| 3.4 | 4 | 2 | 0 | 0 | 0 | 3.4 | 6 | 4 | 0 | 0 | 0 |
| 3.5 | 2 | 1 | 0 | **0** | 0 | 3.5 | 3 | 0 | 0 | **0** | 0 |
| 3.6 | 1 | 0 | 0 | 0 | 0 | 3.6 | 5 | 0 | 0 | 0 | 0 |
| 3.7 | 0 | 0 | 0 | 0 | 0 | 3.7 | 0 | 0 | 0 | 0 | 0 |
| 3.8 | 0 | 0 | 0 | 0 | 0 | 3.8 | 0 | 0 | 0 | 0 | 0 |
| 3.9 | 0 | 0 | 0 | 0 | 0 | 3.9 | 0 | 0 | 0 | 0 | 0 |
| | False, Nass <15 | | | | | | False, Nass <15 | | | | |
| 3.2 | 37 | 16 | 7 | 5 | 2 | 3.2 | 40 | 16 | 10 | 4 | 1 |
| 3.3 | 18 | 8 | 3 | 2 | 0 | 3.3 | 24 | 9 | 3 | 0 | 0 |
| 3.4 | 3 | 2 | 0 | 0 | 0 | 3.4 | 5 | 4 | 0 | 0 | 0 |
| 3.5 | 2 | 1 | 0 | **0** | 0 | 3.5 | 3 | 0 | 0 | **0** | 0 |
| 3.6 | 1 | 0 | 0 | 0 | 0 | 3.6 | 5 | 0 | 0 | 0 | 0 |
| 3.7 | 0 | 0 | 0 | 0 | 0 | 3.7 | 0 | 0 | 0 | 0 | 0 |
| 3.8 | 0 | 0 | 0 | 0 | 0 | 3.8 | 0 | 0 | 0 | 0 | 0 |
| 3.9 | 0 | 0 | 0 | 0 | 0 | 3.9 | 0 | 0 | 0 | 0 | 0 |

Table 10. The average spacing between subsequent arrivals for one template as a function of STA and SNRcc threshold for two time segments with the highest aftershock activity as observed between 3 a.m. and 10 a.m. on September 3, 2017 (7 hours) and between 10 a.m. and 12 a.m. on November 3, 2020 (2 hours).

| 2017246 (4-10) | STA,s | | | | | 2020308 (10-12) | STA,s | | | | |
|---|---|---|---|---|---|---|---|---|---|---|---|
| SNRcc threshold | 0.5 | 0.6 | 0.7 | 0.8 | 0.9 | | 0.5 | 0.6 | 0.7 | 0.8 | 0.9 |
| 3.2 | 109 | 138 | 164 | 183 | 211 | 3.2 | 112 | 137 | 164 | 187 | 220 |
| 3.3 | 131 | 171 | 209 | 237 | 273 | 3.3 | 133 | 169 | 206 | 234 | 288 |
| 3.4 | 161 | 218 | 274 | 310 | 361 | 3.4 | 165 | 216 | 266 | 306 | 377 |
| 3.5 | 204 | 282 | 359 | **410** | 480 | 3.5 | 209 | 286 | 360 | **408** | 506 |
| 3.6 | 264 | 368 | 487 | 545 | 634 | 3.6 | 263 | 373 | 484 | 534 | 696 |
| 3.7 | 342 | 489 | 646 | 720 | 824 | 3.7 | 341 | 502 | 631 | 714 | 859 |
| 3.8 | 448 | 636 | 838 | 924 | 1041 | 3.8 | 441 | 647 | 808 | 926 | 1094 |
| 3.9 | 589 | 812 | 1091 | 1149 | 1295 | 3.9 | 566 | 791 | 1052 | 1134 | 1368 |

For another two days without aftershocks, the estimates of Δt are presented in Table 11, and Table 12 lists the statistics of false event hypotheses. A conservative set of control parameters for September 4, 2017, is STA=0.9 s and threshold of 3.5, and for August 30, 2017, 0.8 s and 3.5, respectively. The estimates of ∆t are ~420 s and ~550 s for 2017242 and 2017247, respectively. The rate of false events on September 4, 2017, is higher than on August 30 might be related to the presence of weak signals coherent with the DPRK templates. No aftershocks can be detected at regional distances, but one can expect some aftershock activity one day after the mainshock, which can be only detected at local distances. The empirical ∆t value for 2017242 is compatible with the theoretical estimates. This observation evidences in favor of the assumption on a random and independent distribution of detections during the days without aftershocks.

Table 11. The average spacing between subsequent arrivals for one template as a function of STA and SNRcc threshold for two full days (25 hours) without aftershocks found in this study. The day 2017242 is four days before and 2017247 is the next day after the main shock.

| 2017242 | STA,s | | | | | 2017247 | STA,s | | | | |
|---|---|---|---|---|---|---|---|---|---|---|---|
| SNRcc threshold | 0.5 | 0.6 | 0.7 | 0.8 | 0.9 | | 0.5 | 0.6 | 0.7 | 0.8 | 0.9 |
| 3.2 | 112 | 143 | 172 | 197 | 231 | 3.2 | 112 | 132 | 171 | 192 | 225 |
| 3.3 | 136 | 182 | 224 | 257 | 313 | 3.3 | 136 | 183 | 223 | 255 | 303 |
| 3.4 | 170 | 239 | 304 | 355 | 427 | 3.4 | 170 | 241 | 292 | 333 | 399 |
| 3.5 | 221 | 318 | **418** | 490 | 590 | 3.5 | 222 | 323 | 404 | 458 | **553** |
| 3.6 | 292 | 431 | 582 | 686 | 836 | 3.6 | 295 | 443 | 565 | 638 | 770 |
| 3.7 | 393 | 614 | 815 | 958 | 1123 | 3.7 | 401 | 587 | 788 | 886 | 1051 |
| 3.8 | 530 | 852 | 1142 | 1321 | 1534 | 3.8 | 553 | 807 | 1108 | 1228 | 1448 |
| 3.9 | 732 | 1191 | 1508 | 1785 | 2059 | 3.9 | 748 | 1107 | 1544 | 1689 | 1962 |

Tuning of the WCC control parameters is based on the statistics of the true/false detections and event hypotheses obtained for different periods of aftershock activity within the DPRK test site. For the periods with known aftershocks found by the WCC method and in IDC routine processing, one has two targets: 1) to build as reliable event hypotheses as possible; 2) retain the rate of false event hypotheses at the level of ~1 per month. The first target suggests the set of control parameters providing the highest resolution and sensitivity. The second target requires a more conservative approach. In order to resolve this conflict, we adopted a two-stage approach. The set for the routine WCC processing is optimal for poor hypotheses' suppression: STA=0.8 s, SNRcc threshold 3.5, CWL=120 s, filters from #2 to #4. These parameters are practically optimal for the detection of the weakest aftershocks and save computation resources. For example, Figure 12 illustrates the number of detections in the association intervals in the routine WCC processing for 03.09.2017. There are no intervals which can potentially generate a false even hypothesis. This is an evidence of the efficient suppression of false events. The case of 12.03.2011 in Figure 12 demonstrates that the highest possible rate of signals generated by many sources at regional distances from stations KSRS/USRK and the DPRK test site practically does not affect the creation of event hypotheses related to the DPRK aftershocks.

The low resolution is somehow compensated by the reduced Nass=11. The event hypotheses with 11≤Nass≤19 are not considered reliable enough for the final XSEL. The second stage is needed to take the final decision on these seed events. A similar approach is adopted at the IDC: the result of automatic processing, SEL3, is checked by IDC analysts for potential improvement to the REB level or rejection. For the WCC method, the set with the highest sensitivity and resolution is used: STA=0.5 s and SNRcc threshold of 3.5. A lower threshold may generate an enormous number of false detections potentially suppressing valid detections with resulting a reduction in Nass instead of improvement.

Table 12. The number of false events created by cross-correlation as a function of STA and SNRcc threshold for two full days (25 hours) without aftershocks found in this study. The day 2017242 is four days before and 2017247 is the next day after the mainshock. False events have two categories: 11≤Nass<15, considered as fully random or related to side sensitivity, and 15≤Nass≤19, considered as potentially valid events for further study.

| 2017242 | STA, s | | | | | 2017247 | STA, s | | | | |
|---|---|---|---|---|---|---|---|---|---|---|---|
| | False, Nass <20 | | | | | | False, Nass <20 | | | | |
| SNRcc threshold | 0.5 | 0.6 | 0.7 | 0.8 | 0.9 | | 0.5 | 0.6 | 0.7 | 0.8 | 0.9 |
| 3.2 | 44 | 10 | 7 | 2 | 1 | 3.2 | 22 | 18 | 7 | 2 | 1 |
| 3.3 | 8 | 5 | 2 | 1 | 0 | 3.3 | 21 | 4 | 4 | 1 | 0 |
| 3.4 | 4 | 0 | 0 | 0 | 0 | 3.4 | 4 | 1 | 1 | 0 | 0 |
| 3.5 | 2 | 0 | **0** | 0 | 0 | 3.5 | 2 | 0 | 0 | 0 | **0** |
| 3.6 | 0 | 0 | 0 | 0 | 0 | 3.6 | 0 | 0 | 0 | 0 | 0 |
| 3.7 | 0 | 0 | 0 | 0 | 0 | 3.7 | 0 | 0 | 0 | 0 | 0 |
| 3.8 | 0 | 0 | 0 | 0 | 0 | 3.8 | 0 | 0 | 0 | 0 | 0 |
| 3.9 | 0 | 0 | 0 | 0 | 0 | 3.9 | 0 | 0 | 0 | 0 | 0 |
| | False, Nass <15 | | | | | | False, Nass <15 | | | | |
| 3.2 | 43 | 10 | 7 | 2 | 1 | 3.2 | 21 | 18 | 7 | 2 | 1 |
| 3.3 | 8 | 5 | 2 | 1 | 0 | 3.3 | 21 | 3 | 4 | 1 | 0 |
| 3.4 | 4 | 0 | 0 | 0 | 0 | 3.4 | 4 | 0 | 1 | 0 | 0 |
| 3.5 | 2 | 0 | **0** | 0 | 0 | 3.5 | 0 | 0 | 0 | 0 | **0** |
| 3.6 | 0 | 0 | 0 | 0 | 0 | 3.6 | 0 | 0 | 0 | 0 | 0 |
| 3.7 | 0 | 0 | 0 | 0 | 0 | 3.7 | 0 | 0 | 0 | 0 | 0 |
| 3.8 | 0 | 0 | 0 | 0 | 0 | 3.8 | 0 | 0 | 0 | 0 | 0 |
| 3.9 | 0 | 0 | 0 | 0 | 0 | 3.9 | 0 | 0 | 0 | 0 | 0 |

Figure 20 illustrates the importance of the second high-resolution processing stage. The routine WCC processing of 2020308 has generated a list of detections. There were association intervals with NDT≥20, 15≤NDT<20, and 11≤NDT<15. There were also several intervals with 6≤NDT<11, which do not initiate the Local Association procedure but prove the absence of event hypotheses except those related to NDT≥11. In the routine WCC processing, two XSEL events were built, and two more aftershocks were found in the high-resolution processing, which was applied to the time intervals with Nass≥15.

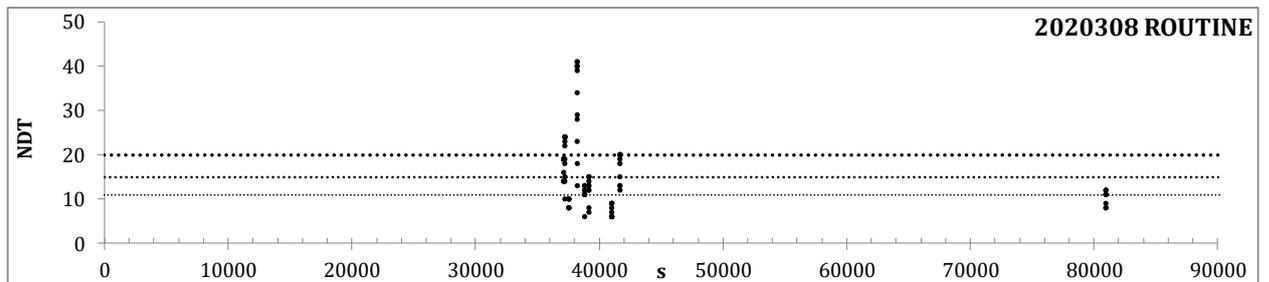

Figure 20. The number of detections, NDT, in the running eight-second-intervals with 1 s step as estimated for 03.11,2020.

The control parameters of the WCC processing were tuned to find all reliable DPRK aftershocks and to reject all unreliable event hypotheses. At the routine stage, the STA=0.8 s, the SNRcc threshold is set to 3.5, the number of associated templates Nass≥11, the LTA=120 s, the CWL is fixed to 120 s, the spacing between subsequent detections by the same master is 60 s, only filters

#2 through #4 are used. An event hypothesis has to match all EDCs in the automatic regime, but rule 6) is applied in the interactive regime. There are more than 70 event hypotheses after 11.09.2016 found using this setting. There are several hypotheses which were found in the high-resolution processing with STA=0.5 s, SNRcc threshold 3.6, the number of associated templates Nass≥15, LTA=120 s, the CWL varies from 20 s to 120 s with a 20 s step, the spacing between subsequent detections by the same master is 30 s, all five filters used. The high-resolution mode is applied only to the one-hour-long intervals with the event hypotheses obtained in the routine processing. The final solution on the event hypotheses obtained in both regimes to be promoted to the XSEL is also made in the interactive regime. The XSEL includes only events with Nass≥20, but some exclusions are also possible.

For the periods without known aftershocks, one can apply the optimal settings for the routine stage. This approach is used to process the period after September 9, 2016, as we searched for reliable aftershocks. Currently, routine WCC processing aimed at the DPRK aftershocks also uses the same set of control parameters and detected many reliable aftershocks in 2021, which are also reprocessed with the high-resolution set [Kitov, 2021]. The period between 01.01.2009 and 09.09.2016 has to be processed with the optimal sensitivity and resolution in order to find the aftershock hypotheses similar to those detected between 11.09.2016 and the current date. The number of aftershocks has been growing since the DPRK5 explosion, but we did not find any unusual events so far. There were some attempts to process short periods (few weeks) after the DPRK2, DPRK3, and DPRK at the initial stage of the aftershock analysis, with lower sensitivity and resolution, however. No aftershocks were found before the first one occurred on 11.09.2016 [Adushkin *et al*., 2017]. A comprehensive reprocessing with the well-calibrated multi-master method is needed.

**7. Results of the two-stage WCC processing between 01.01.2009 and 09.09.2016**

We have processed the whole period before the DPRK5 with a conservative set of control parameters and found several event hypotheses matching the EDCs adopted in this study. Table 13 lists these solutions together with a few hypotheses formally not matching the EDCs but considered as good candidates for the second processing stage. There are 8 reliable events with Nass>20. These aftershocks are very similar to those detected after the DPRK5 and DPRK6 in the shape of signals, as confirmed by the SNRcc values in Table 14, as well as by the *RM* and Nass. Four of the eight aftershocks (2013249, 2014145_2, 2014145_3, and 2016184) are in the top 25% of all aftershocks by Nass, which is the principal parameter defining their reliability. The event 2014145_3 is the largest among all found aftershocks with Nass=48. We interpret the largest events as related to the cavity collapse: the events on 25.05.2014 (2014145) and on 02.07.2016 (2016184) may manifest the final stage of this process since no other similar events were observed thereafter.

The supposedly false event hypotheses with Nass<15 were rejected, but their number (<100 between 01.01.2009 and 09.09.2016) and origin represent a good estimate for the assessment of the routine processing resolution. The statistics of event hypotheses with 11≤Nass<15 is important for the understanding of the variation in the generation of false events below the decision line separating the conditionally false events used as seeds for the high-resolution stage and the unconditionally false events. A larger part of the 11≤Nass<20 hypotheses were related to the side lobes of the CC-detector and were rejected by the EDC #6, *i.e.* by the *RM* mismatch. These events are not counted as false because they are not random, and their sources are clear. They were rejected in the interactive regime and not used for the statistics. For example, on 11.03.2011 the largest rate of such high-magnitude events per day is observed. The event on 04.09.2016 (2016250) was selected for the second stage because of its closeness to the DPRK5 despite it has Nass=13. This event was not improved in the high-resolution processing to the level of the XSEL, and thus, was rejected.

The event found on 19.02.2016 (2016050) is of special interest. There are some indications that there was a radioactive signal detected approximately 6 weeks after the DPRK4 (06.01.2016). There is no direct connection between these two observations but their synchronized occurrence is of potential interest for further investigation. An aftershock may manifest the processes leading to the radioactive release or can be related to the final stage of cavity collapse when all radioactive gases already left the cavity. The 2016050 aftershock was weak and hardly related to the final cavity collapse. It is important that this weak event was found in the routine WCC processing and improved at the high-resolution processing stage from Nass=24 to Nass=30. Six associated templates added to the event hypothesis significantly strengthen the final solution and promote this aftershock up in the hierarchy of reliability. This is the effect expected from the second stage. As an alternative, it can be no link between the RN gas release and aftershocks. For example, the first DPRK3 aftershock was found on 29.05.2013 (2013149) and the radioactive gas release was detected by the IMS RN-stations in April 2013.

Table 13. Aftershocks of the DPRK3 and DPRK4 found in the routine WCC processing. The final solutions obtained in the high resolution processing are also presented.

| Date | Time | Routine | | High-resolution | | | |
|---|---|---|---|---|---|---|---|
| | | Nass | RM | Nass | KSRS | USRK | RM |
| 2013149_1 | 13:53:08 | 34 | 2.63 | 40 | 16 | 24 | 2.65 |
| 2013150_1 | 02:28:23 | 15 | 2.88 | 20 | 5 | 15 | 2.87 |
| 2013200_1 | 13:23:50 | 25 | 2.61 | 28 | 10 | 18 | 2.63 |
| 2013203_1 | 15:50:16 | 23 | 2.34 | 22 | 6 | 16 | 2.40 |
| 2014145_1 | 04:56:56 | 25 | 2.69 | 26 | 8 | 18 | 2.69 |
| 2014145_2 | 06:01:27 | 40 | 2.79 | 43 | 18 | 25 | 2.82 |
| **2014145_3** | 06:43:03 | 48 | 2.94 | 48 | 23 | 25 | 2.91 |
| 2016050_1 | 00:28:07 | 24 | 2.78 | 30 | 6 | 24 | 2.80 |
| 2016090_1 | 02:17:13 | 17 | 3.15 | 20 | 10 | 10 | 3.14 |
| 2016112_1 | 12:39:52 | 16 | 3.16 | 15 | 10 | 6 | 3.12 |
| **2016184_1** | 19:52:27 | 36 | 2.61 | 38 | 19 | 19 | 2.61 |
| 2016250_1 | 06:02:03 | 13 | 2.75 | 15 | 4 | 11 | 2.76 |

One of the most important observations is that there were no aftershocks of the DPRK2 and there were no other events similar to the DPRK aftershocks between 01.01.2009 and 12.02.2013. More than four years were processed with the conservative set of control parameters and no events, even with Nass≥15, were detected. This is strong evidence in favor of the assumption that there was no natural seismicity near the DPRK test site which can be detected by the most sensitive multi-master method. At the same time, after the DPRK3 and DPRK4, we detected several events similar to the aftershocks of the DPRK5 and DPRK6. The conservative set of control parameters at the routine stage may suppress some weak aftershocks with signals hidden in the ambient noise. The chance to miss a weak DRPK2 aftershock, as well as weak aftershocks of the DRPK3 and the DPRK4, is a challenge for a comprehensive study of the DPRK aftershock activity and we address this issue by processing longer intervals of several months with a set of control parameters close to the high-resolution set.

The level of similarity between the pre-DPRK5 aftershocks and the waveform templates used in the multi-master method can be illustrated by the SNRcc values. Table 14 presents the maximum

SNRcc within ±3 s from the empirical (or predicted if a template is not associated with the event) arrival time for all templates at stations KSRS and USRK. (For each aftershock, we use the color scale to highlight the SNRcc value from the largest (green) and lowest (red) values separately for the templates of explosions, from #1 to #7, and aftershocks.) For all new aftershocks, the highest SNRcc is above 10 for all templates at USRK and almost all at KSRS. Figures 13 through 16 demonstrate that the number of SNRcc above 10 is low in the absence of real aftershocks. The largest SNRcc among all new aftershocks is 26.9 at KSRS and 24.1 at USRK, both belong to the event 2014145_3. These values prove a very high level of cross-correlation. One can suggest that the pre-DPRK5 seismic events are of the same nature as the DPRK aftershocks induced by the DPRK5 and DRPK6 underground tests.

The evolution of the aftershock activity between 01.01.2009 and 09.09.2016 is presented in Figure 21. There are two different measures used to illustrate the aftershock properties: *RM* and Nass. Three DPRK underground nuclear tests conducted within the studied period are also presented in order to illustrate the link between these tests and their aftershocks. There are no events found between DPRK2 and DPRK3 - more than 3 years. The events related to the DPRK3 and DPRK4 show different patterns: it took more than 100 days to the first DPRK3 aftershock and ~40 days for the DPRK4. The biggest DPRK3 aftershock occurred ~470 days after the mainshock. For the DPRK4, the same elapsed time falls into the period of DPRK5 activity. No large events are observed before the DPRK6 except the DPRK5 aftershock on 11.09.2016. One cannot exclude the possibility that one of the DPRK6 aftershocks is actually the collapse of the DPRK4 cavity.

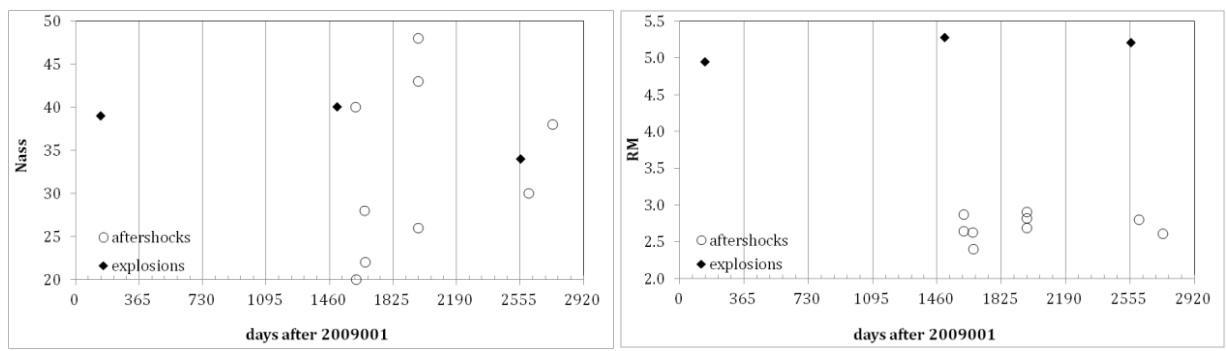

Figure 21. The evolution of the DPRK aftershock activity between January 1, 2009, and September 9, 2016. Two measures are presented: *RM* and Nass. The DPRK underground tests are also shown to illustrate the link between the aftershocks and the tests. Vertical lines have a spacing of 365 days.

The second stage of the WCC processing used the set of parameters allowing the highest resolution and sensitivity without an extraordinary number of false events, however. Table 13 lists the results of this processing in the hour-long intervals, which included the event hypotheses obtained in the routine stage. Only the successful examples are shown and a few representative failures. Two events could be added: 2016090_1, which was promoted from Nass=17 to Nass=20, and 2013150_1 with five new templates associated. The event 2016090_1 raises some doubts, however, because of *RM*=3.14. It was removed from the list of true event hypotheses. Two events: 2016112_1 and 2016250_1 were not promoted to the XSEL according to the Nass<20 rule. No other events within the studied period had Nass>15 to be processed in the second stage.

Almost all true event hypotheses were improved at the second stage. Only the event 2013203_1 has lost one associated template. The event 2013149_1 obtained 6 new associated templates. Other events have marginal or no improvement. In Figure 22 the time distribution of SNRcc for detections at KSRS and USRK is presented for the high-resolution setting. The upper panel displays a one-hour window which includes the largest DPRK3 aftershocks on 25.05.2014. In the lower panel, the case of 02.07.2016 is presented. All three solutions are reliable.

Table 14. Max SNRcc values estimated within ±3 s from the empirical arrival times for the associated templates and the predicted ones, if the template is not associated, at stations KSRS and USRK.

| Date_order# | 2013149_1 | 2013150_1 | 2013200_1 | 2013203_1 | 2014145_1 | 2014145_2 | 2014145_3 | 2016050_1 | 2016184_1 |
|---|---|---|---|---|---|---|---|---|---|
| Master | | | | | KSRS | | | | |
| 3875968 | 2.9 | 2.6 | 2.7 | 2.2 | 2.0 | 3.2 | 4.2 | 3.0 | 8.4 |
| 5397597 | 4.2 | 2.7 | 3.5 | 2.2 | 2.8 | 3.9 | 6.4 | 2.7 | 8.2 |
| 9486691 | 4.3 | 3.6 | 2.5 | 3.0 | 2.6 | 4.0 | 6.6 | 2.5 | 6.5 |
| 12733679 | 4.7 | 2.8 | 3.2 | 3.0 | 2.9 | 4.7 | 7.5 | 3.2 | 6.0 |
| 13552792 | 4.7 | 2.7 | 2.9 | 2.7 | 2.8 | 4.3 | 7.2 | 2.5 | 7.6 |
| 14812788 | 5.1 | 2.9 | 3.1 | 2.8 | 2.9 | 4.8 | 7.5 | 3.1 | 6.7 |
| 14807656 | 3.2 | 3.1 | 2.7 | 3.0 | 2.5 | 3.8 | 6.2 | 3.0 | 4.6 |
| 20172661 | 4.0 | 2.9 | 5.2 | 3.4 | 3.1 | 5.5 | 8.3 | 2.7 | 17.7 |
| 20172662 | 4.8 | 2.4 | 6.0 | 4.2 | 4.1 | 8.0 | 12.0 | 2.8 | 20.4 |
| 20172851 | 3.5 | 2.6 | 5.3 | 4.1 | 2.8 | 6.0 | 7.4 | 3.4 | 18.7 |
| 20173041 | 2.4 | 3.1 | 2.7 | 2.5 | 2.0 | 2.6 | 2.9 | 1.9 | 1.9 |
| 20173351 | 2.9 | 2.6 | 2.8 | 2.8 | 2.2 | 4.4 | 4.3 | 2.0 | 10.1 |
| 20173391 | 3.0 | 2.4 | 3.9 | 3.6 | 2.4 | 5.7 | 6.8 | 3.2 | 14.2 |
| 20173401 | 2.9 | 2.1 | 2.6 | 2.2 | 2.8 | 2.5 | 3.3 | 2.5 | 4.1 |
| 20173431 | 2.4 | 2.9 | 2.4 | 2.8 | 2.9 | 2.5 | 2.7 | 2.9 | 4.2 |
| 20173432 | 2.3 | 1.8 | 4.5 | 3.2 | 2.9 | 4.0 | 4.8 | 2.9 | 7.5 |
| 20173433 | 2.2 | 2.7 | 2.8 | 2.3 | 2.9 | 2.4 | 3.7 | 3.0 | 2.7 |
| 20162551 | 4.3 | 2.7 | 3.0 | 2.2 | 4.0 | 6.9 | 11.5 | 3.9 | 4.5 |
| 20172462 | 4.0 | 3.0 | 2.4 | 2.4 | 3.1 | 5.9 | 10.2 | 2.7 | 3.8 |
| 20180361 | 3.6 | 2.2 | 2.1 | 2.4 | 3.4 | 7.4 | 9.8 | 3.7 | 3.0 |
| 20180362 | 3.5 | 2.6 | 2.2 | 2.0 | 3.3 | 6.9 | 8.3 | 2.7 | 3.2 |
| 20180363 | 3.2 | 2.3 | 2.7 | 2.3 | 2.8 | 5.0 | 6.8 | 2.6 | 2.5 |
| 20180371 | 2.3 | 2.1 | 2.3 | 2.8 | 2.7 | 3.1 | 4.4 | 1.9 | 2.4 |
| 20180372 | 1.9 | 3.0 | 2.7 | 1.9 | 3.2 | 3.5 | 5.5 | 2.5 | 1.9 |
| 20180373 | 11.0 | 4.3 | 5.7 | 4.2 | 8.4 | 18.9 | 25.5 | 6.5 | 7.1 |
| 20180381 | 8.9 | 4.6 | 5.1 | 4.5 | 7.7 | 18.4 | 22.6 | 4.4 | 9.1 |
| 20180391 | 3.3 | 2.9 | 3.1 | 1.9 | 3.6 | 8.3 | 11.1 | 3.4 | 3.4 |
| 20181121 | 5.9 | 4.1 | 3.4 | 2.6 | 4.4 | 8.7 | 14.1 | 5.3 | 3.3 |
| 20181122 | 11.5 | 5.2 | 6.0 | 4.1 | 8.7 | 20.7 | 26.9 | 6.1 | 7.5 |
| | | | | | USRK | | | | |
| 5397597 | 5.6 | 4.0 | 5.4 | 4.1 | 4.6 | 7.8 | 11.8 | 5.2 | 7.5 |
| 9486691 | 5.6 | 3.6 | 6.5 | 3.3 | 5.1 | 8.0 | 10.6 | 3.8 | 7.6 |
| 12733679 | 7.2 | 3.4 | 4.0 | 4.7 | 4.6 | 7.4 | 9.2 | 5.7 | 5.1 |
| 13552792 | 5.5 | 3.6 | 6.2 | 4.9 | 4.3 | 6.3 | 9.9 | 4.5 | 6.5 |
| 14812788 | 5.5 | 3.4 | 4.0 | 3.3 | 4.3 | 7.8 | 9.0 | 5.0 | 4.1 |
| 14807656 | 5.1 | 4.7 | 4.6 | 4.3 | 4.2 | 6.4 | 8.0 | 4.4 | 4.2 |
| 20172661 | 10.3 | 5.2 | 5.6 | 5.8 | 3.7 | 7.2 | 8.9 | 4.1 | 6.3 |
| 20172662 | 11.5 | 7.2 | 10.4 | 10.8 | 6.2 | 10.2 | 16.4 | 10.4 | 13.8 |
| 20172851 | 11.3 | 6.5 | 9.2 | 6.5 | 5.9 | 11.2 | 16.5 | 7.9 | 10.2 |
| 20173041 | 2.3 | 2.7 | 2.6 | 2.6 | 2.7 | 3.2 | 4.5 | 3.0 | 2.7 |
| 20173351 | 6.7 | 5.0 | 6.8 | 5.7 | 4.3 | 7.8 | 11.7 | 4.5 | 7.7 |
| 20173391 | 8.9 | 4.1 | 7.3 | 6.0 | 5.3 | 10.1 | 16.1 | 7.3 | 8.3 |
| 20173401 | 5.2 | 3.5 | 3.2 | 2.8 | 3.3 | 4.9 | 5.7 | 4.4 | 4.1 |
| 20173431 | 5.2 | 3.2 | 2.5 | 2.8 | 3.5 | 3.6 | 5.0 | 4.4 | 4.6 |
| 20173432 | 8.7 | 3.7 | 3.8 | 4.9 | 4.5 | 5.3 | 7.1 | 7.7 | 6.8 |
| 20173433 | 9.0 | 3.6 | 4.1 | 5.1 | 5.1 | 5.7 | 8.6 | 7.9 | 7.1 |
| 20162551 | 4.3 | 3.1 | 2.9 | 3.3 | 3.6 | 6.1 | 6.3 | 3.9 | 2.3 |
| 20172462 | 3.5 | 2.6 | 4.0 | 3.2 | 3.0 | 3.5 | 5.0 | 3.6 | 3.0 |
| 20180361 | 8.0 | 6.3 | 6.3 | 6.8 | 8.7 | 15.1 | 18.2 | 7.0 | 4.0 |
| 20180362 | 4.5 | 3.1 | 3.8 | 3.0 | 6.0 | 4.6 | 8.7 | 4.4 | 2.5 |
| 20180363 | 2.4 | 3.4 | 2.0 | 2.5 | 4.2 | 6.7 | 8.3 | 3.6 | 2.5 |
| 20180371 | 2.8 | 3.6 | 2.1 | 3.3 | 4.1 | 7.6 | 7.6 | 2.6 | 2.0 |
| 20180372 | 3.5 | 3.9 | 2.5 | 2.8 | 4.9 | 8.6 | 9.6 | 3.2 | 2.4 |
| 20180373 | 11.2 | 7.2 | 7.6 | 6.9 | 9.1 | 15.0 | 23.3 | 8.9 | 4.8 |
| 20180381 | 8.8 | 6.1 | 7.2 | 5.9 | 7.8 | 14.7 | 20.0 | 7.0 | 3.9 |
| 20180391 | 5.2 | 4.4 | 3.3 | 4.2 | 5.6 | 9.1 | 9.6 | 5.6 | 2.3 |
| 20181121 | 8.5 | 5.4 | 4.0 | 4.2 | 6.3 | 8.5 | 14.8 | 6.9 | 3.7 |
| 20181122 | 12.0 | 8.6 | 6.6 | 6.2 | 10.3 | 18.1 | 24.1 | 8.3 | 4.9 |

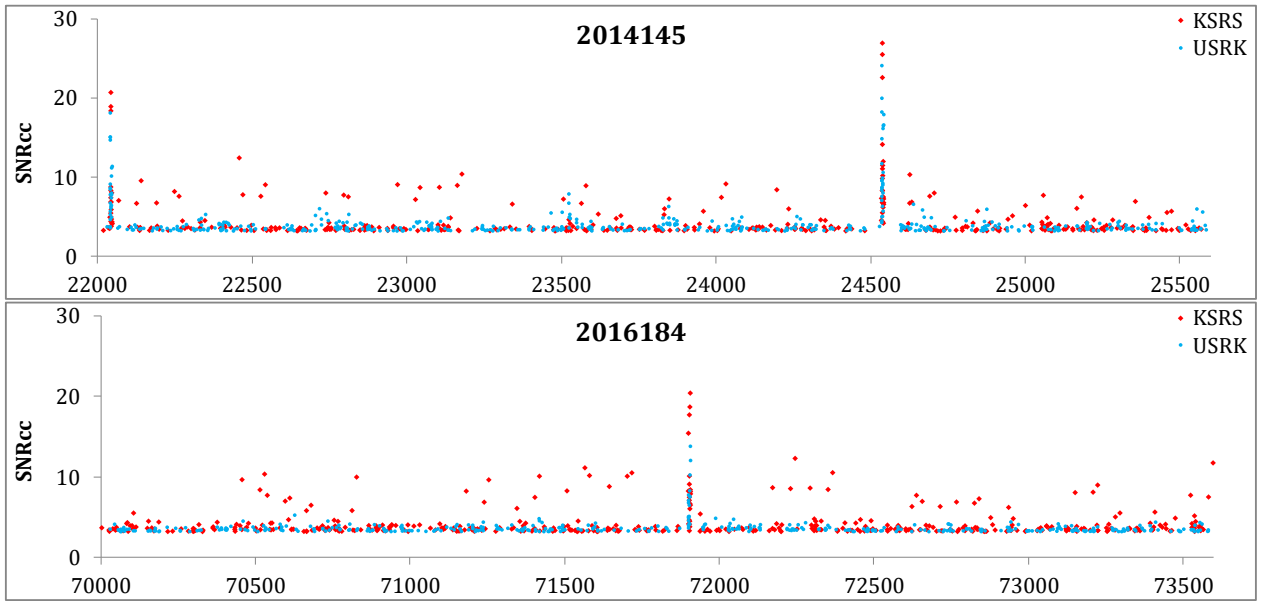

Figure 22. One-hour intervals which include the aftershocks potentially related to the DPRK3 (upper panel) and DPRK4 (lower panel) cavity collapses.

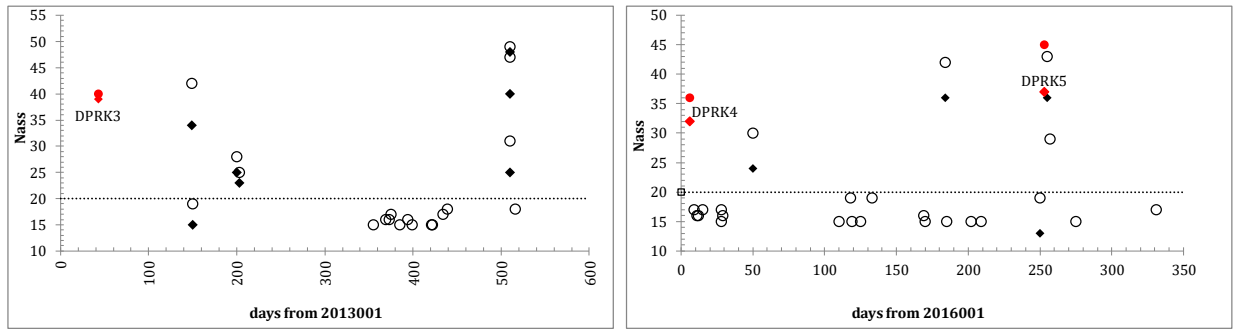

Figure 23. The result of the WCC processing of longer periods after DPRK3 and DPRK4 using an extended set of control parameters close to those in the high-resolution mode. Red symbols - explosions. Solid diamonds - the routine solutions. Open circles - the intermediate regime solutions.

There is a question, however, what could we find if the high-resolution mode is applied to the daily data? Instead of direct usage of the high-resolution setting with a high rate of false event hypotheses, we have created a slightly reduced set of control parameters: STA=0.5 s, SNRcc threshold 3.6, all 5 filters, CWL from 40 s to 120 s with a 40 s step. The short CWL is a potential source of false detections, and the choice of longer windows may suppress the rate of false events. The 0.1 increase in the detection threshold also helps to remove numerous false detection. At the same time, this set of parameters has a higher resolution than that in the routine regime. Figure 23 presents the result of processing in this intermediate regime. There were no new valid event hypotheses in the DPRK3 sequence, but the solutions in the intermediate regime demonstrate significant improvement on the routine solutions. The same observation is valid for the DPRK4 intermediate-resolution processing: no new events added despite many seed events are added with Nass<20. The processing is limited to 350 days and includes the after DPRK5 period. For the DPRK2, we processed the days between 2009140 and 2009365 as well as the first 100 days in 2010. Several events with Nass>15 and *RM*>4 were rejected. There were 6 event hypotheses with Nass≥15 during this period. None of them became valid event hypotheses after the high-resolution processing and we do not show the period after the DPRK2 in Figure 23. These results suggest that there were no aftershocks of the DPRK2 which can be detected by the multi-master method at regional stations KSRS and USRK. No additional aftershocks were found after the DPRK3 and DPRK4.

The routine and high-resolution stages of WCC processing were successful in finding valid events and suppression of false event hypotheses. The routine stage did not generate many event hypotheses due to the conservative approach to detection and local association. Since the high-resolution set was applied only to specific hours one might express some doubts about the absence of very weak events missed in routine processing. To address these doubts, we have processed the years between 2009 and 2017 using a set of WCC control parameters similar to the high-resolution mode. The only important difference is the SNRcc threshold of 3.6 instead of 3.5 in the high-resolution mode and a different set of CWLs. There were no aftershocks found in the intermediate mode extra to those detected in the routine WCC processing. Overall, seven DPRK3 and two DPRK4 aftershocks were found. They were not reported before, and thus, provide significant information on the DPRK tests. These events were found by the multi-master method, whose sensitivity and resolution depend on the aftershocks of the DPRK5 and DPRK6. Without these aftershocks, the task to find the DPRK3 and DPRK4 aftershocks would be a difficult challenge.

## 8. Discussion

The aftershock activity of the DPRK5 and DPRK6 initiated a series of studies related to their nature and potential usage for monitoring of the CTBT [Adushkin *et al*., 2017; Kim *et al.*, 2018; Schaff *et al*., 2018; Tian *et al*., 2018]. The aftershock sequence has been developing in time and demonstrates higher activity in 2021 than in 2019. The depth of burial of the DPRK5 and DPRK6 as well as the close proximity of their epicenters may explain the mechanical causes of the observed acceleration in the number of aftershocks in 2021 [Adushkin *et al*., 2021; Kitov, 2021]. The growing attention to the current state of the DPRK5 and DPRK6 cavities and collapse chimneys has a negative impact on the study of potential aftershock activity related to the earlier tests: from the DPRK2 to the DPRK4.

The study of the DPRK5/DPRK6 aftershocks allowed us to develop and test the multi-master method with high sensitivity and resolution. The use of this method allowed to find 90+ aftershocks since 09.09.2016, some of them extremely weak as measured by magnitude of 2.0 to 2.5. It was methodologically correct to apply the multi-master method to the data available since January 1, 2009, and prove the absence of natural seismicity within the DPRK test site at least at the level of the method resolution. Several attempts to find aftershocks of the DPRK2, DPRK3, and DPRK4 failed because of the low sensitivity of the old methods and the absence of appropriate waveform templates.

The result of continuous WCC processing of the data from IMS array stations KSRS and USRK brought a few surprises - the aftershocks of the DPRK3 and DPRK4. There were no events found between 01.01.2009 and 12.02.2013, *i.e.* between the DPRK2 and DPRK3. This observation supports the assumption of no natural events within the DPRK test site: there are no signals having a good correlation with the waveform templates from the DPRK5 and DPRK6 aftershocks during the period where aftershocks are absent. The surprise aftershocks were weak or far enough in time from the explosions that induced them. To some extent, it explains the failure of the previous aftershock search covered a few weeks after the events. Such short periods of processing were driven by the observation of the DPRK5 and DPRK6 aftershocks just in a few days after the mainshock. The 1.5-year delay between the explosion and the cavity collapse is an unusual observation [Kitov and Kuznetsov, 1990; Adushkin and Spivak, 1993].

Table 15. Max SNRcc values estimated within ±3 s from the empirical arrival times for the associated templates and the predicted ones, if the template is not associated, at stations KSRS and USRK for DPRK nuclear tests. The color scale is event-specific.

| Date_order# | 2009145_0 | 2013043_0 | 2016006_0 | 2016253_0 | 2017246_0 | 2009145_0 | 2013043_0 | 2016006_0 | 2016253_0 | 2017246_0 |
|---|---|---|---|---|---|---|---|---|---|---|
| Master | KSRS | | | | | USRK | | | | |
| 3875968 | 26.1 | 20.4 | 17.9 | 24.0 | 17.6 | | | | | |
| 5397597 | | 67.2 | 44.6 | 64.9 | 45.2 | | 68.3 | 44.0 | 57.4 | 43.4 |
| 9486691 | 68.2 | | 56.7 | 63.4 | 54.6 | 74.7 | | 53.3 | 62.3 | 51.3 |
| 12733679 | 45.4 | 51.8 | | 68.1 | 72.9 | 42.7 | 50.8 | | 64.8 | 57.6 |
| 13552792 | 65.0 | 60.8 | 71.4 | | 68.3 | 51.9 | 51.3 | 59.0 | | 61.9 |
| 14812788 | 52.0 | 56.7 | 80.9 | 77.5 | | 50.7 | 54.0 | 65.5 | 68.4 | |
| 14807656 | 20.7 | 22.0 | 24.5 | 22.9 | 28.9 | 9.9 | 12.0 | 21.5 | 28.3 | 26.2 |
| 20172661 | 8.7 | 11.2 | 5.9 | 7.1 | 7.9 | 6.1 | 5.7 | 5.3 | 6.2 | 7.0 |
| 20172662 | 12.6 | 16.1 | 11.1 | 14.9 | 16.5 | 8.0 | 8.5 | 6.2 | 6.7 | 8.4 |
| 20172851 | 7.9 | 9.5 | 7.1 | 8.7 | 8.8 | 7.7 | 6.8 | 5.8 | 6.9 | 6.1 |
| 20173041 | 3.2 | 3.3 | 3.5 | 4.1 | 4.0 | 3.4 | 3.7 | 4.2 | 3.9 | 3.7 |
| 20173351 | 4.2 | 4.7 | 4.0 | 4.4 | 4.2 | 5.2 | 3.7 | 4.0 | 4.5 | 3.1 |
| 20173391 | 6.4 | 5.9 | 5.6 | 5.9 | 6.9 | 6.7 | 5.2 | 6.2 | 6.6 | 6.7 |
| 20173401 | 4.3 | 4.3 | 4.4 | 4.6 | 5.1 | 3.6 | 3.7 | 3.6 | 3.8 | 3.9 |
| 20173431 | 4.0 | 3.0 | 2.8 | 3.0 | 2.6 | 3.3 | 3.0 | 3.1 | 3.1 | 2.5 |
| 20173432 | 9.0 | 8.2 | 5.3 | 5.1 | 3.1 | 6.4 | 7.0 | 6.1 | 5.5 | 5.5 |
| 20173433 | 4.6 | 4.5 | 4.9 | 4.4 | 2.5 | 5.1 | 5.2 | 5.1 | 5.2 | 4.9 |
| 20162551 | 4.3 | 6.1 | 4.7 | 4.8 | 4.1 | 8.0 | 8.7 | 10.5 | 12.1 | 12.7 |
| 20172462 | 4.1 | 4.5 | 4.0 | 3.8 | 3.5 | 5.1 | 5.4 | 5.7 | 6.1 | 7.0 |
| 20180361 | 2.8 | 2.8 | 2.4 | 2.5 | 2.6 | 6.1 | 6.9 | 5.8 | 4.8 | 4.6 |
| 20180362 | 3.0 | 3.0 | 2.5 | 3.4 | 3.1 | 4.7 | 4.4 | 3.3 | 3.2 | 2.4 |
| 20180363 | 2.8 | 2.7 | 2.3 | 2.2 | 2.4 | 3.8 | 3.8 | 3.3 | 3.0 | 3.4 |
| 20180371 | 2.2 | 2.4 | 2.1 | 2.6 | 2.4 | 3.6 | 4.7 | 3.8 | 4.0 | 4.4 |
| 20180372 | 2.6 | 2.9 | 2.4 | 2.8 | 2.5 | 3.8 | 4.6 | 4.1 | 3.9 | 4.7 |
| 20180373 | 7.4 | 7.7 | 6.1 | 7.2 | 7.4 | 10.0 | 10.6 | 10.3 | 9.8 | 9.0 |
| 20180381 | 9.2 | 9.9 | 9.9 | 11.1 | 11.0 | 11.5 | 12.8 | 11.8 | 9.8 | 10.7 |
| 20180391 | 3.3 | 3.5 | 3.1 | 3.0 | 3.3 | 3.8 | 4.9 | 5.5 | 4.7 | 4.4 |
| 20181121 | 3.9 | 4.0 | 3.3 | 3.7 | 3.3 | 6.6 | 6.6 | 4.8 | 5.3 | 5.0 |
| 20181122 | 6.4 | 7.4 | 6.7 | 7.1 | 7.6 | 9.8 | 10.4 | 9.4 | 9.5 | 10.6 |

Several weak events not found in routine IDC processing and also missed by the cross-correlation methods raise a principal question on their nature. Could it be a clandestine nuclear test with an extremely low yield? As an alternative, it could be a partial fission reaction, a cavity decoupled explosion [Adushkin *et al*., 1992] or a chemical explosion. The date of the biggest DPRK3 aftershock on 25.05.2014 is approximately in the middle between the DPRK3 and DPRK4. There are two measures one can use to answer this challenge. The P/S spectral ratio of the DPRK aftershocks detected after the DPRK5 is an excellent discrimination criterion [Kitov *et al*., 2017]. For the 2014145 as well as for the 2016184 events, which we interpret as related to cavity collapse processes, the spectral ratio belongs to the population of the DRPK aftershocks [Kitov *et al*., 2018]. For smaller pre-DPRK5 events, the P/S spectral ratio is not reliably estimated since the corresponding signals are close to the ambient noise. Another way to compare the DPRK explosions and the DPRK3/DPRK4 aftershocks is to compare the SNRcc values obtained for the explosion and aftershock waveform templates. Table 14 lists the SNRcc for the pre-DPRK5 aftershocks and Table 15 presents similar estimates for five DPRK tests, DPRK1 is excluded as no data at USRK was available. A simple comparison of the SNRcc estimates for four ME/SE combinations in Tables 14 and 15 shows that the templates from the DPRK5 and DPRK6 aftershocks have a much higher correlation with the signals from the pre-DRPK5 aftershocks than with the signals from the DPRK tests. The explosion-related templates have an extremely high correlation with the explosion signals and a lower correlation with the

pre-DPRK5 events found in this study. Therefore, cross-correlation supports the assumption that the DPRK3 and DPRK4 events are of the same nature as the DPRK5 and DPRK6 aftershocks.

Further interpretation of the DPRK3 and DPRK4 aftershocks together with later aftershocks of the DPRK5 and DPRK6 has many aspects. The depth of burial is of principal importance for the understanding of the gravitational energy that can be released in seismic events related to cavity collapse. The absolute moment tensor and relative moment tensors for the DPRK explosions with aftershocks may shed some light on the difference in correlation between signals generated by the observed aftershocks. The cross-correlation parameters like *CC* and SNRcc are good indicators of similarity. The relative magnitude is rather satiated for very small events, and other measures of the relative size are needed [Kitov *et al*., 2014].

There is an important problem of relative location. The multi-master method is based on weak aftershocks, and the corresponding estimates of the arrival times are not accurate enough to achieve the same accuracy in relative location as for the DPRK underground tests. All these problems were not addressed in this study in full. We focused on detection and local association problems and found several reliable aftershocks missed in the previous studies. At the same time, we proved that there were no other events in the zone around the DPRK test site which can be interpreted as not the DPRK aftershocks, at least at the level of the multi-master sensitivity and resolution. Data from seismic stations much closer than KSRS and USRK would be helpful to confirm the DPRK aftershocks found in this study and also to test the hypothesis of the natural seismic events' absence around the test site.